\documentclass{aa}
\usepackage[varg]{txfonts}
\usepackage{epstopdf}
\usepackage{graphicx}
\usepackage{subfig}
\usepackage{natbib}
\bibpunct{(}{)}{;}{a}{}{,}
\usepackage{longtable}
\usepackage{array}
\usepackage{sidecap}
\usepackage{lscape}
\usepackage{color}
\usepackage{amstext}

\begin{document}

 \title{Chromospheric activity catalogue of 4454 cool stars.}
 \subtitle{Questioning the active branch of stellar activity cycles}
\author{S.~Boro Saikia\inst{\ref{inst1}, \ref{inst2}}, C.~J.~Marvin\inst{\ref{inst1}}, S.~V.~Jeffers\inst{\ref{inst1}}, A.~Reiners\inst{\ref{inst1}}, R.~Cameron\inst{\ref{inst3}}, S.~C.~Marsden\inst{\ref{inst4}}, P.~Petit\inst{\ref{inst5},
\ref{inst6}}, J.~Warnecke\inst{\ref{inst3}}, and  A.~P.~Yadav\inst{\ref{inst1}}
}
\institute{Institut f\"ur Astrophysik, Georg-August-Universit\"at G\"ottingen, Friedrich Hund Platz 1, 37077 G\"ottingen, Germany\label{inst1}
\and Institut f\"ur Astrophysik, Universit\"at Wien, T\"urkenschanzstrasse 17, A-1180 Vienna, Austria \label{inst2}
 \and Max-Planck-Institut f\"ur Sonnensystemforschung, Justus-von-Liebig-Weg 3, 37077 G\"ottingen, Germany\label{inst3}
\and University of Southern Queensland, Computational Engineering and Science Research Centre, Toowoomba 4350, Australia \label{inst4}
\and CNRS, Institut de Recherche en Astrophysique et Plan\'etologie, 14 Avenue Edouard Belin, F-31400 Toulouse, France \label{inst5}
\and Universit\'e de Toulouse, UPS-OMP, Institut de Recherche en Astrophysique et Plan\'etologie, Toulouse, France \label{inst6}
}
\authorrunning{S.~Boro Saikia, C.~J.~Marvin et al.}
\abstract
{{}Chromospheric activity monitoring of a wide range of cool stars can provide valuable information on stellar magnetic activity and its dependence on fundamental stellar parameters such as effective temperature and rotation.}
{We compile a chromospheric activity catalogue of 4454 cool stars from a combination of archival HARPS spectra and multiple other surveys, including the
Mount Wilson data that have recently been released by the NSO. We explore the variation in chromospheric activity of 
cool stars along the main sequence for stars with different effective temperatures. Additionally, we also perform an activity-cycle period 
search and investigate its relation with rotation.}
{The chromospheric activity index, S-index, was measured for 304 main-sequence stars from archived high-resolution HARPS spectra. 
Additionally, the measured and archived S-indices were converted into the chromospheric flux ratio log $R'_\mathrm{HK}$.
The activity-cycle periods were determined using the generalised Lomb-Scargle periodogram to study the active and inactive branches on 
the rotation \textendash \ activity-cycle period plane.} 
{The global sample shows that the bimodality of chromospheric activity, known as
the Vaughan-Preston gap, is not prominent, with a significant percentage of the stars at an intermediate-activity level around $\log R'_\mathrm{HK}$ = -4.75. 
Independently, the cycle period search shows that stars can lie in the region intermediate between the active and inactive branch, 
which means that the active branch is not as clearly distinct as previously thought.}
{The weakening of the Vaughan-Preston gap indicates that cool stars spin down from a higher activity level and settle at a 
lower activity level without a sudden break at intermediate activity.
Some cycle periods are close to the solar value between the active and inactive branch, 
which suggests that the solar dynamo is most likely a common case of the stellar dynamo.}
\maketitle 
\section{Introduction}
The non-thermal emission in the line cores of certain chromospheric lines such as  the \ion{Ca}{II} H\&K, \ion{Mg}{II}, H$\alpha$, and \ion{Ca}{II} {infrared triplet} is widely used as an indicator of the surface magnetic activity in cool stars (spanning F to M dwarfs). The first successful measurement of 
emission in stellar chromospheric lines was carried out by \citet{eberhard13}. Arguably the most renowned study of stellar chromospheric emission is the 
Mount Wilson program, which measured the chromospheric activity of more than a thousand stars over a time span of more than four decades \citep{wilson68,
duncan91,baliunas95}. The Mount Wilson survey indicates that stars can be categorised into either active or inactive stars, with a lack of stars with intermediate 
activity.
\paragraph{}
The lack of stars in the intermediate-activity region is known as the Vaughan-Preston gap \citep{vaughanpreston80}. Different theories have been proposed to 
explain this gap. Firstly, it could be an indication of two distinct physical phenomena, such as different stellar populations \citep{baliunas95,henry96}, 
stellar modes or topologies \citep{durney81,middlekoop82,vitense07}, or rapid shifts in differential rotation \citep{metcalfe16}. Secondly, it could be a 
temporal effect, caused by a smooth \citep{noyes84}, rapid \citep{pace09}, or critical spin-down rate of the rotation \citep{vaughanpreston80,middlekoop82,
noyes84}. Lastly, as suggested by \citet{vaughanpreston80}, it could be just a matter of small number statistics, and the Vaughan-Preston gap might not be real
\citep{vaughanpreston80,noyes84}.
\paragraph{}
In addition to measuring chromospheric activity for hundreds of stars, the Mount Wilson project also established that cool stars other than the Sun have magnetic activity
cycles. With observations spanning several decades, the chromospheric activity cycles of the Mount Wilson sample were first 
investigated by \citet{baliunas95}. These authors reported three different types of activity cycles: solar-like cyclic activity, highly 
variable non-cyclic activity, and flat activity. Observations of cyclic chromospheric activity in the Mount Wilson sample \citep{baliunas95} suggest that 
solar-like chromospheric activity is common and is exhibited by many cool dwarfs. 
\paragraph{}
The detection of magnetic activity cycles in solar-like stars provides valuable observational constraints on solar and stellar dynamo models. The mechanism 
behind the dynamo-driven activity cycle was investigated by many groups \citep{noyes84a,baliunas85,brandenburg98,saar99,vitense07}. For old stars with 
clear cyclic behaviour, \citet{noyes84a} reported a possible correlation between activity-cycle period $P_\mathrm{cyc}$ and Rossby number, Ro, where Ro is 
the ratio of the rotation period to the convective overturn time ($P_\mathrm{rot}/\tau_\mathrm{c}$). However, \citet{baliunas85} did not find any correlation 
between activity-cycle period $P_\mathrm{cyc}$, rotation period $P_\mathrm{rot}$, Ro, and other stellar properties. 
\paragraph{}
To investigate the relationship between stellar activity cycles and the dynamo process, the well-constrained activity cycles in the Mount Wilson sample
were investigated by \citet{saar92}. They reported that stellar activity cycles form two distinct branches when plotted as a function of rotation. 
They classified them as  the active and inactive branch. The active branch comprises stars with strong activity, and the inactive branch consists of 
stars with weak chromospheric activity. These two branches were also confirmed by \citet{brandenburg98}, \citet{saar99} and \citet{brandenburg17}. Both 
\citet{brandenburg98} and \citet{saar99} investigated the dependence of the $\alpha$ effect on the magnetic field ($B$) for stars with a known chromospheric 
activity cycle. According to dynamo models, the $\alpha$ effect describes the inductive effect of a twisted magnetic field via turbulent convective helical motions
that can generate a poloidal from a toroidal magnetic field \citep{steenbeck66}. Together with the $\Omega$ effect, which is differential rotation, it has been 
successful in reproducing different aspects of the solar magnetic cycle under the mean-field dynamo models \citep[see e.g.][for a detailed review on the 
solar dynamo]{ossendrijver03,charbonneau10}. Despite the advances made in numerical modelling, no current dynamo model can fully reproduce all observed solar 
phenomena.
\paragraph{}
To provide observational constraints for current dynamo models, \citet{brandenburg98} and \citet{saar99} used the ratio of activity cycle and rotation frequency 
($\omega_\mathrm{cyc}/\Omega=P_\mathrm{rot}/P_\mathrm{cyc}$) and defined Ro as $1/2\Omega\tau_\mathrm{c}=P_\mathrm{rot}/4\pi\tau_\mathrm{c}$. In this 
definition, $\omega_\mathrm{cyc}/\Omega$ is proportional to the $\alpha$ effect when we assume that $\alpha$ and the radial shear are linearly proportional to 
$\Omega$. One key result was that stars in both branches show evidence of a dynamo, where $\alpha$ increases linearly with the magnetic field strength $B$, 
and not, as predicted, by simple $\alpha$ quenching $\alpha\propto B^{-2}$. \citet{saar99} also reported the presence of multiple cycles, where stars from 
the active branch could have a second cycle on the inactive branch, and vice versa. However, they also noted that the classification of stars into two branches
based on the activity and the evolution of the cycles from one branch to another might be more complex than the clear distinction reported in their work. 
Multiple cycles and migration of stars from one branch to another was also reported by \citet{vitense07} in a separate model-independent 
study that investigated cycle period versus rotation period. Recently, \citet{metcalfe16} updated the figure of \citet{vitense07} showing rotation versus cycle period with approximately 30 stars. \citet{metcalfe16} concluded that the Sun is an anomaly amongst other Mount Wilson stars with similar rotation periods. It is the only star 
with an activity cycle that does not lie on either the inactive or active branch. However, the sample size is small and the results might be different if 
we were to include the large Mount Wilson sample that has recently been
 released by the National Solar Observatory (NSO)  and stars from other long-term surveys. 
\footnote{http://www.nso.edu/node/1335}
\paragraph{}
We here compile a chromospheric activity catalogue of cool stars. We achieve this by combining archival data sets available in \textit{Vizier} with data from the Mount  Wilson survey \citep{duncan91,baliunas95} (released by the NSO)\footnote{A sub sample of the NSO data, comprising of 
solar analogues, was also investigated by \cite{egeland17}} and the HARPS F, G, K, and M radial velocity targets \citep{lovis11,bonfils13}. The Mount 
Wilson data released by the NSO also consist of S-index measurements that were not published in \citet{baliunas95}, with some observations made until 2000. 
We analyse the full Mount Wilson sample for the first time since \citet{duncan91} and \citet{baliunas95} and investigate the average and the long-term activity of cool 
stars along the main sequence. 
\paragraph{}
One key motivation for compiling the catalogue is to monitor the chromospheric activity of cool stars as a function of effective temperature, or $B-V$. 
We investigate if there is indeed a lack of intermediately active cool stars in the solar neighbourhood. Additionally, we explore trends in stellar activity 
cycles as a function of rotation to provide observational constraints to solar and stellar dynamo models. This paper is organised as follows: In Section 2 
we describe the archival data and introduce the catalogue. In Section 3 we discuss the chromospheric activity in terms of the S-index and the $R'_\mathrm{HK}$
conversion for the global sample. For a sub-sample of stars with long-term measurements, the activity-cycle periods are calculated in Section 4. The relation 
between activity-cycle period and rotation is investigated in Section 5, followed by a comparison with Zeeman Doppler imaging results in Section 6, and 
we finally provide our conclusions in Section 7.
\section{Data collection}
\begin{figure}
\centering
\includegraphics[scale=0.5]{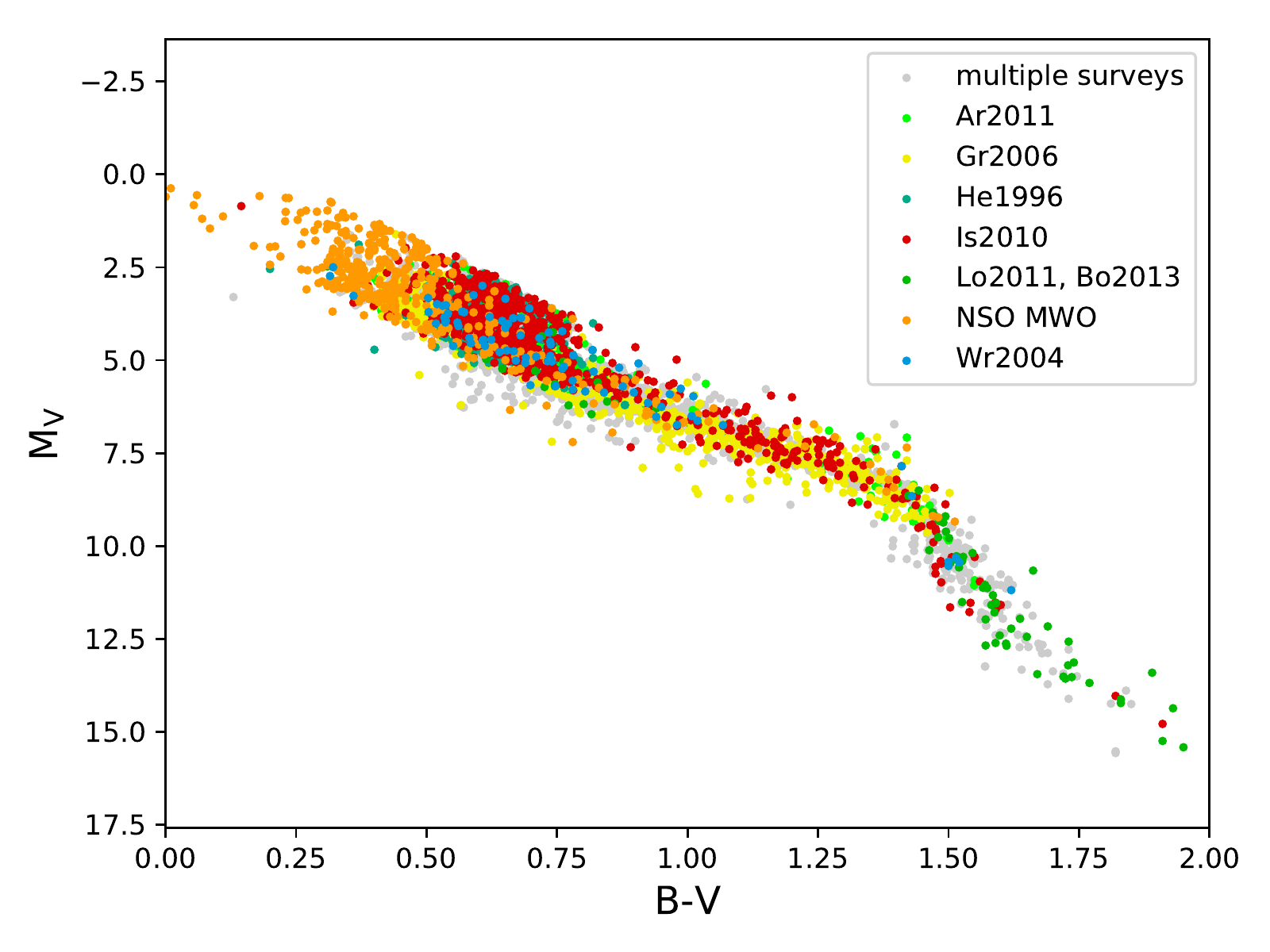}
\caption{HR diagram of the stars in the catalogue. The labels represent the different surveys (see Section 2 for more details on the surveys). The stars that appear in multiple surveys are also indicated with a different colour.}
\label{HR}
\end{figure}
The total number of unique cool stars (main-sequence dwarfs) in the catalogue is 4454, 2509 of which can be found in single surveys, and 1945 appear in 
multiple surveys. The Hertzsprung-Russel (HR) diagram of the cool stars in the catalogue is shown in Fig. \ref{HR}. Except for the HARPS stars, all other S-index measurements 
were taken from existing values in the literature. 
\paragraph{}
The surveys also contain several evolved stars that we have removed in this analysis. 
We used the main-sequence $B-V$-$M_V$ relation from Table B.1 in \citet{Gray2005book} and removed stars lying outside the relation with $M_V \pm 1$. 
The number of cool stars corresponding to different surveys including references is listed in Table \ref{data}. The stellar properties of the cool dwarfs 
are listed in Table \ref{maintable}. Only a few of the stars are shown here. The full catalogue will be available in \textit{Vizier}.
A detailed description of the sources is given below.
\begin{itemize}
\item Ar2011: The Magellan survey \citep{arriagada11}
provided 643 stars. For individual stars, the median S-index ($S_\mathrm{med}$) was calibrated to 
the Mount Wilson scale. 
\item Wr2004 and Is2010: The California and Carnegie planet search
provided 1101 stars \citep{wright04}, 739 of which have both mean 
($S_\mathrm{mean}$) and median ($S_\mathrm{med}$) S-index values, and 362 stars have only $S_\mathrm{med}$ values. Additionally, 1798 stars were also 
observed as part of the California planet survey \citep{isaacson10};
they have only $S_\mathrm{med}$ values. 
\item Gr2006 and He1996: A survey of southern solar-type stars carried out by \citet{henry96} and \citet{gray06}. The southern sample of stars in \citet{henry96} consists
of 758 stars within 50 pc of the solar neighbourhood. For the stars that were observed several times, $S_\mathrm{mean}$ was taken from the literature. 
$S_\mathrm{mean}$ of 1288 southern stars was also taken from \citet{gray06}. The southern stars only have $S_\mathrm{mean}$ published in the literature. 
\item Ha2009: $S_\mathrm{med}$ of 28 stars were taken from the solar and stellar activity program of the
Lowell observatory \citep{hall09}. 
\item Lo2011 and Bo2013: High-resolution archival spectra were obtained for 304 F, G, K stars and 103 M dwarfs \citep{lovis11, bonfils13}.  We analysed the 
spectra of 304 F, G, K stars and calculated the S-index calibrated to the Mount Wilson S-index based on the technique used by Marvin et al. 2018 (in prep; hereafter M18; for a brief description of the method, we refer to Section 3). The M-dwarf spectra were analysed by M18. 
\item NSO MWO: Data from the Mount Wilson survey \citep{duncan91,baliunas95} have recently been re-released by the NSO and are included in the catalogue. We included 827 cool 
stars from the NSO Mount Wilson survey. The median and mean S-indices were calculated from the long-term measurements. 
\end{itemize}
When an S-index measurement was found in several surveys, the median and mean were calculated from multiple measurements. The median was preferred over the mean 
because when it is observed over long-term flares, it might influence the mean S-index. A summary of the stars included in the catalogue and their sources is listed 
in Table \ref{data}.
\begin{table*}
\centering
\caption{Archival surveys and the number of stars corresponding to each survey, including the references. The surveys are listed in column 2, the total 
number of main-sequence and evolved stars are listed in column 3, the total number of main-sequence stars is presented in column 4, the mean $S_\mathrm{MWO}$ is listed in column 5, the median
$S_\mathrm{MWO}$ is given in column 6, and the references are
collected in column 7.}
\label{data}
\begin{tabular}{cccccccc}
\hline
\hline
No&survey&Total no.&No. of &$S_\mathrm{mean}$&$S_\mathrm{med}$&Reference\\
&&of stars&MS stars&&&\\
\hline
(i)  & Magellan       & 673  & 634  & No  & Yes & 1\\ %
(ii) & CPS            & 808  & 739  & Yes & Yes & 2\\ %
      &                & 401  & 362  & Yes & No  & 2\\ %
      &                & 2620 & 1798 & No  & Yes & 3\\ %
(iii)  & Southern stars & 825  & 758  & Yes & No  & 4\\ %
      &                & 1359 & 1288 & Yes & No  & 5\\ %
(iv)   & Lowell         & 28   & 28   & Yes & Yes & 6\\ %
(v)  & HARPS          & 311  & 304  & Yes & Yes & 7\\ %
      &                & 116  & 103  & Yes & Yes & 8\\ %
(vi) & NSO MW 1995    & 1903 & 827  & Yes & Yes & 9\\ %
\hline
 Total & (w/o duplicates) &   &  4454 &    &     & \\ %
\end{tabular}
\tablebib{
    (1)~\citet{arriagada11};(2)~\citet{wright04}; (3)~\citet{isaacson10}; (4)~\citet{henry96}; (5)~\citet{gray06}; (6)~\citet{hall09}; (7)~\citet{lovis11}; (8)~\citet{bonfils13}; (9)~\citet{baliunas95} and \citet{duncan91}, released by the NSO
 in 2016.
See Section 2 for more details on the type of S-index measurements.
}
\end{table*}
\begin{longtab}
\begin{landscape}
\begin{longtable}{lcccccccccccccc}
\caption{\label{maintable}Chromospheric activity and stellar parameters for stars in the catalogue. The column headers are the name, spectral type, $B-V$,
RA, DEC, absolute V magnitude, parallax, mean S-index ($S_\mathrm{mean}$), median S-index ($S_\mathrm{med}$), the standard deviation in S-index 
($S_\mathrm{std}$), and $\log R'_\mathrm{HK}$. The stellar parameters were taken from HIPPARCOS unless otherwise stated. Only a 
portion of the table is shown here. The full catalogue will be available in $Vizier$.}\\
\hline\hline
name&spectype&$B-V$&RA(ICRS)&DE(ICRS)&Mv&Plx&$S_\mathrm{mean}$&$S_\mathrm{med}$&$S_\mathrm{std}$&$\log R'_\mathrm{HK}$\\
\hline
HD102117&G6V&0.721&176.21055164&-58.70353973&4.35379697702&23.81&0.105&0.106&0.006&-5.006\\
HD115585&G5/G6IV/V&0.742&199.96369673&-70.85494655&4.24335464863&23.05&0.101&0.102&0.003&-5.073\\
HD12345&G8III&0.746&30.2025882&-12.87526774&5.64288478528&23.8&0.118&0.120&0.005&-4.910\\
HD134985&K1&0.772&228.19897523&-1.16540113&6.21427764592&24.37&0.131&0.131&0.006&-4.828\\
HD162236&G9V&0.726&267.78070309&-27.39610817&5.71993914874&23.12&0.19&0.19&0.01&-4.58\\
HD166724&K0IV/V&0.861&273.49852847&-42.57518256&6.16304855196&23.26&0.22&0.22&0.02&-4.63\\
HD16714&G5V&0.708&39.94819506&-33.89908714&5.28709271176&24.97&0.124&0.124&0.001&-4.87\\
HD203384&K0&0.762&320.47312814&-4.17399146&5.03872541119&25.22&0.116&0.117&0.003&-4.977\\
HD223121&K1V&0.94&356.73211614&-31.98167606&6.29355871469&24.25&0.18&0.17&0.02&-4.78\\
HD25673&K0&0.816&61.08444054&-4.65558342&6.45176707069&24.23&0.14&0.14&0.02&-4.83\\
HD31822&G0&0.581&74.64910667&-9.33405665&4.76087816385&23.13&0.129&0.129&0.001&-4.77\\
HD34449&G3V&0.614&72.63706679&-85.37507484&4.96743992445&23.2&0.131&0.131&0.002&-4.747\\
HD3569&K0V&0.846&9.63821988&-12.17262517&6.11286501533&24.02&0.149&0.147&0.007&-4.851\\
HD6348&G5&0.801&16.11080178&-2.36626356&6.18099711832&25.48&0.141&0.143&0.007&-4.814\\
HD86171&G5&0.746&149.16074258&-8.83486532&5.70124726198&27.18&0.171&0.169&0.008&-4.672\\
HD967&G5&0.645&3.5176132&-11.31111824&5.23190849025&23.68&0.128&0.128&0.003&-4.791\\
HD98356&K0V&0.828&169.66610892&-10.12638938&5.59823736105&23.64&0.176&0.175&0.004&-4.757\\
HIP473&K6V+M0.5V&1.41&1.41782275&45.81245496&7.84964780042&85.1&1.65&1.64&0.07&-4.52\\
HIP795&G6V+G7V&0.716&2.46506329&8.45319673&4.40297038964&14.21&0.45&0.45&0.02&-4.33\\
HIP4849&K3V&1.008&15.60155392&5.0609096&6.50239551466&46.61&0.50&0.51&0.02&-4.61\\
HIP5881&G5&0.671&18.88025814&37.74350026&5.00943352307&16.9&0.20&0.20&0.02&-4.76\\
HIP7080&G5&0.76&22.80805715&-10.89662492&4.22273943841&17.32&0.145&0.145&0.001&-5.115\\
SAO11844&tmp&0.52&23.7811&60.7816&3.49114354298&&0.146&0.145&0.004&-4.92\\
GJ3126&tmp&1.5&30.3917&63.77&10.4215803134&78.4&1.11&1.13&0.06&-4.82\\
HIP9724&M2.5&1.514&31.26683543&-17.61420901&10.315299843&105.94&1.2&1.2&0.1&-4.8\\
HIP10492&G6IV&0.737&33.81735061&-23.28140736&4.88429651479&23.06&0.225&0.222&0.009&-4.753\\
HIP11843&F8V&0.572&38.22562452&15.03446457&3.71039073168&34.84&0.18&0.18&0.01&-4.76\\
HIP12764&F5&0.534&41.00359859&-6.00937192&3.85299717862&22.73&0.157&0.156&0.002&-4.859\\
HIP13027&G1V+G5V&0.68&41.86395248&19.37221052&4.33286075259&30.66&0.39&0.38&0.01&-4.37\\
HIP19786&G0&0.64&63.61327492&12.43534652&4.78078651116&22.19&0.31&0.31&0.01&-4.45\\
HIP20899&G2V&0.609&67.20096632&17.28553719&4.4503828987&21.09&0.31&0.32&0.01&-4.42\\
HIP22715&K3V&1.019&73.2693207&22.23548477&6.62628416664&37.09&0.643&0.643&0.004&-4.496\\
HIP23452&K7V&1.43&75.61879215&-21.25610576&8.65797052589&117.38&1.4&1.39&0.04&-4.62\\
HIP23786&G9V&0.804&76.6751978&14.44681583&5.84068027487&41.7&0.32&0.32&0.01&-4.61\\
HIP23932&M3.5V&1.52&77.14473389&-18.1686451&10.4329986098&107.3&0.97&0.94&0.06&-4.91\\
HIP25662&G0V&0.582&82.21491883&12.55135347&4.40224025855&34.55&0.162&0.161&0.003&-4.862\\
SAO170732&tmp&0.36&85.8373&-20.1893&3.26759215189&&0.239&0.237&0.003&-4.446\\
HIP31083&G0&0.71&97.83981164&2.911269&4.90455725108&35.72&0.177&0.175&0.006&-4.9\\
HIP35519&K0&0.876&109.97365142&9.22967182&5.70570741725&19.64&0.175&0.175&0.005&-5.073\\
HIP38657&K2.5V&0.95&118.7250404&19.23744468&6.25702040908&50.05&0.186&0.187&0.002&-5.103\\
\hline
\hline
\end{longtable}
\end{landscape}
\end{longtab}
\section{Chromospheric activity}
The chromospheric activity of cool stars can be determined by measuring the flux in the chromospheric \ion{Ca}{II} H\&K lines and normalising it 
to the nearby continuum, which is  commonly referred to as the S-index \citep{vaughanpreston80,duncan91}.
\subsection{S-index}
The majority of {S-index values included in this catalogue were taken from} archives where the value is already calibrated to the Mount Wilson scale. We denote 
the S-index on the Mount Wilson scale as $S_\mathrm{MWO}$. We calculated the S-index for the archival HARPS spectra and calibrated it to $S_\mathrm{MWO}$.
\subsubsection{Archival S-index values} 
While some surveys consist of both mean and median S-index values, only the mean or median $S_\mathrm{MWO}$ is published for some surveys. We combined these 
archived values with the measured $S_\mathrm{MWO}$ from HARPS to create the global sample. 
\subsubsection{S-index calculation for HARPS spectra}
For HARPS spectra, we followed the prescription of \citet{duncan91} to mimic the response of the Mount Wilson HKP-2 spectrophotometer. We took a triangular 
bandpass at the core of the H and K lines at 3968.47 $\AA$ and 3933.664 $\AA$, respectively, and two 20 $\AA$ wide rectangular bandpasses to measure the 
nearby continuum; $V$ centred on 3901.07 $\AA$ and $R$ centred on 4001.07 $\AA$. The equation is given as
\begin{equation}
  S_\mathrm{HARPS} = 8 \alpha \frac{N_\mathrm{H} + N_\mathrm{K} }{ N_V + N_R },
  \label{sindex1}
\end{equation}
where $N_\mathrm{H}$, $N_\mathrm{K}$, $N_V$, and $N_R$ are the counts of the respective bandpasses, 8 is a correction factor for the longer exposure times 
of the $V$ and $R$ bandpasses of the HKP-2 instrument, and $\alpha$ is a proportionality constant, usually taken to be $\alpha = 2.4$. We calibrated 
$S_\mathrm{HARPS}$ to $S_\mathrm{MWO}$ by performing a linear regression on stars common in both surveys, so that
\begin{equation}
  S_\mathrm{MWO} = a S_\mathrm{HARPS} + b,
  \label{sharps}
\end{equation}
where $a = 1.1159$ and $b = 0.0343$. The common stars are listed in Table~\ref{table1}. For the HARPS observations, we 
calculated both mean and the median $S_\mathrm{MWO}$.
\begin{figure}
  \centering
  \resizebox{\hsize}{!}{\includegraphics{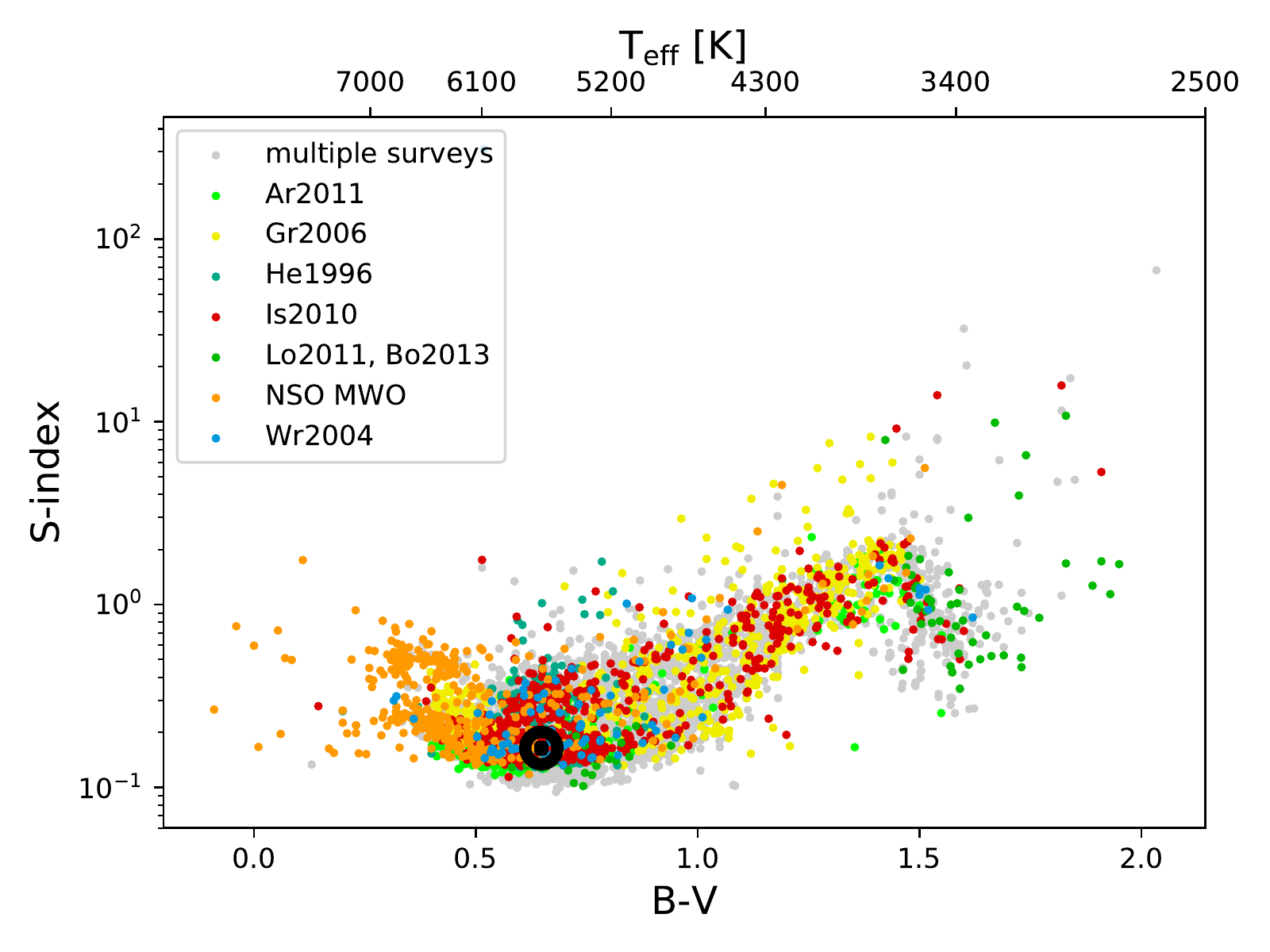}}
  \caption{
    $S_\mathrm{MWO}$ (calibrated to the Mount Wilson scale)
    vs. $B-V$ of {4454} main-sequence stars.
    {The surveys they belong to are given in the legend
    (2509 are only listed in one survey, and 1945 appear in several surveys)}.
    The Sun at minimum activity is shown by the black $\sun$ symbol.
  }
  \label{S}
\end{figure}
\paragraph{}
Figure~\ref{S} shows $S_\mathrm{MWO}$ on a log scale of all available S-index data listed in Table~\ref{data} as a function of $B-V$. We used the median 
$S_\mathrm{MWO}$ when it was available. When $S_\mathrm{med}$ was not available, we used $S_\mathrm{mean}$ . For stars that appeared in several surveys, we plot 
the median $S_\mathrm{MWO}$ value. For $B-V \lesssim 0.93$, the lower activity boundary remains nearly constant. The Sun lies close to the lower activity
boundary. From $0.93 \lesssim B-V \lesssim 1.4$, the lower boundary of $S_\mathrm{MWO}$ increases towards cooler stars. For $B-V \gtrsim 1.40$, the lower 
boundary of activity begins to drop. Over the entire $B-V$ range, the highest activity levels rise towards later spectral types. This colour-dependence 
of the activity is consistent with previous S-index surveys, for example \citet{vaughanpreston80} and \citet{mittag13}. We also plot only the mean or the median, and the 
overall trend remains the same (see Figs. \ref{A1} and \ref{A2} ). 
\subsection{Chromospheric flux ratio log $R'_\mathrm{HK}$}
While $S_\mathrm{MWO}$ gives a measure of \ion{Ca}{II} H and K emission for a relative comparison of stars of similar spectral types, its colour-dependence 
makes it unsuitable for comparing stars of different spectral types. Towards later-type stars, the flux in the $V$ and $R$ channels decreases as the bulk of 
radiation moves towards longer wavelengths. This results in a typical K or M dwarf having a much higher $S_\mathrm{MWO}$ than a typical F or G star. Hence 
we converted the $S_\mathrm{MWO}$ into the chromospheric flux ratio $\log R'_\mathrm{HK}$.
\begin{figure*}
  \centering
  \resizebox{\hsize}{!}{\includegraphics{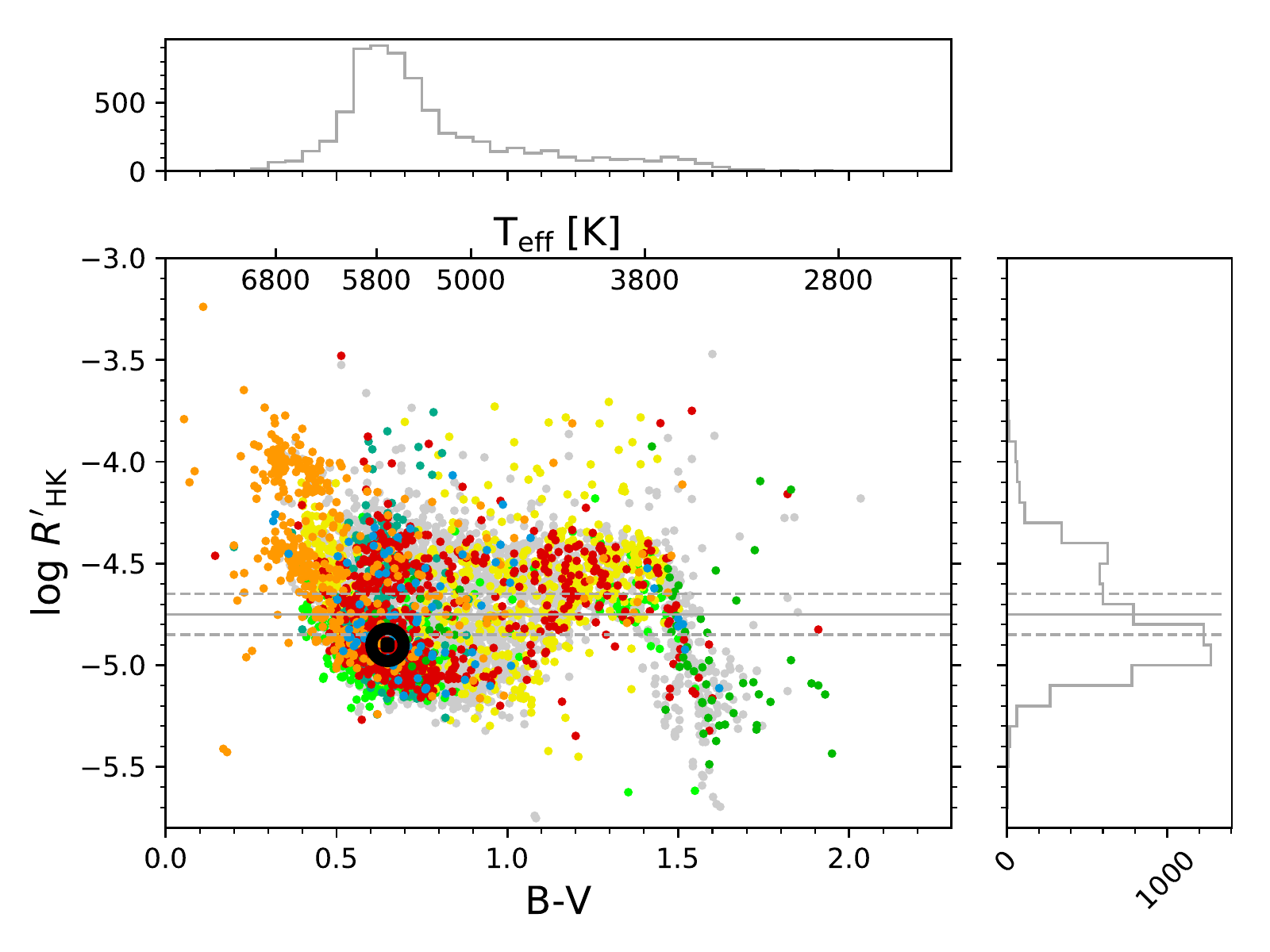}}
  \caption{
    Ratio of chromospheric \ion{Ca}{II} H and K flux
    to bolometric flux, $\log R'_\mathrm{HK}$,
    vs. $B-V$ of {4454} main-sequence stars.
    The symbols and colour palette are the same as in Fig \ref{S}.
    The Sun at minimum activity is shown by the black $\sun$ symbol.
    The dashed grey lines indicate the Vaughan-Preston gap, adopted
from Fig. 2
    of \citet{noyes84}.
    The histogram on the right shows the distribution of chromospheric 
    activity, and the histogram at the top shows the distribution of $B-V$ for stars in the catalogue.
  }
  \label{rphk_plot}
\end{figure*}
\subsubsection{log $R'_\mathrm{HK}$ conversion for archived $S_\mathrm{MWO}$ values}
To compare \ion{Ca}{II} H and K emission of F, G, K, and M dwarfs on the same scale, the ratio of chromospheric flux to bolometric flux is more suitable.
This concept is defined as $R'_\mathrm{HK} = ({F}_\mathrm{HK}-{F}_\mathrm{phot}) / \sigma {T_\mathrm{eff}}^4$ \citep{linsky79} ($F_\mathrm{HK}$ is the flux 
in the H and K lines, $F_\mathrm{phot}$ denotes the photospheric flux and $T_\mathrm{eff}$ stands for the effective temperature). Currently, the most 
frequently used method of measuring $R'_\mathrm{HK}$ is that of \citet{noyes84}, which requires a measurement of $S_\mathrm{MWO}$ and $B-V$ for a particular star. 
The dimensionless $S_\mathrm{MWO}$ is converted into the surface flux $F_\mathrm{HK}$, most commonly by the calibration of \citet{middlekoop82} or the improved 
\citet{rutten84} calibration. The photospheric contribution $R_\mathrm{phot} = {F}_\mathrm{phot}/\sigma {T_\mathrm{eff}^4}$ is most commonly found using 
\citet{noyes84}, who used the \citet{Hartmann84} method. However, this method is valid only for a limited spectral range $0.44 \le B-V \le 0.82$, but many 
stars lie at $B-V > 0.82$.
\paragraph{}
To extend the relation of $R_\mathrm{phot}$ for all the stars in our sample, where $B-V$ can be as high as 2.0, we used the $R_\mathrm{phot}$ polynomial of 
M18. Using PHOENIX stellar atmosphere models, M18 calculated $R_\mathrm{phot}$ values down to $T_\mathrm{eff} = 2300\ \mbox{K}$, which covers the colour 
index range of the stars in our sample. To estimate $T_\mathrm{eff}$, we used the $B-V$ to $T_\mathrm{eff}$ conversion used in \citet{noyes84}. However, the 
computed $R_\mathrm{phot}$ values are higher than those reported
by \citet{noyes84}, as \citet{noyes84} neglected the line core in their calibration, while M18 integrated the 
entire spectral line (see M18 for a more detailed discussion of this). To remain consistent with the relation given by \citet{noyes84}, denoted as 
$\log{R_\mathrm{phot,\,N84}}$, we scaled the relation given by M18, denoted as $\log{R_\mathrm{phot,\,M18}}$, to $\log{R_\mathrm{phot,\,N84}}$:
\begin{equation}
    \log{R_\mathrm{phot,\,N84}} = \log{R_\mathrm{phot,\,M18}} - {0.4612}.
\label{photc}
\end{equation}
The constant {0.4612} was found by a least-squares fit of both relations in the range $0.44 \le B-V \le 0.82$. This correction ensures that stars that lie 
in the \citet{noyes84} calibration range still have similar $R'_\mathrm{HK}$ values to \citet{noyes84}, while extending the relation to later spectral types 
K and M. We also used the $R_\mathrm{HK}$ relation given by M18, which is consistent with both \citet{middlekoop82} and \citet{rutten84}.
\subsubsection{log $R'_\mathrm{HK}$ conversion for HARPS $S_\mathrm{MWO}$}
For HARPS targets, we used the template-model method of M18 to measure $R'_\mathrm{HK}$. This method consists of measuring the surface flux $R_\mathrm{HK}$ by 
co-adding all available spectra into a template spectra with a high signal-to-noise ratio (S/N) using the HARPS-TERRA software \citep{anglada12}. The template spectra were normalised to a 
PHOENIX model atmosphere to convert the arbitrary flux into absolute flux units. Then we integrated the flux in the \ion{Ca}{II} H and K line cores.
The photospheric flux contribution $R_\mathrm{phot}$ was subtracted from $R_\mathrm{HK}$ by subtracting the integrated \ion{Ca}{II} H and K line cores
of the PHOENIX model atmosphere. For consistency, the $R_\mathrm{phot}$ correction in equation \ref{photc} was also applied to the HARPS targets.
\subsubsection{log $R'_\mathrm{HK}$ versus $B-V$}
Figure~\ref{rphk_plot} shows the ratio of chromospheric \ion{Ca}{II} H and K flux to bolometric flux, $\log R'_\mathrm{HK}$ , as a function of $B-V$. Similar 
to Fig. \ref{S}, median values are plotted whenever available, and mean values are used when the median is not available. A total of 4454 stars are plotted, 
with 2509 stars that appear only in one survey and 1945 that appear in several surveys. For stars that appear in several surveys, we plot the median 
$\log R'_\mathrm{HK}$ value. The Sun at minimum activity is shown by a black $\sun$ symbol. The majority of stars lie in the range 
$-5.2 \le \log R'_\mathrm{HK} \le -4.4$. The most active stars have $\log R'_\mathrm{HK} \ge -4.3$ and are located at the top of the plot.
There is a concentration of inactive stars close to the solar activity level of $\log R'_\mathrm{HK} \sim -5.0$. 
\begin{figure*}
\centering
\includegraphics[scale=0.7]{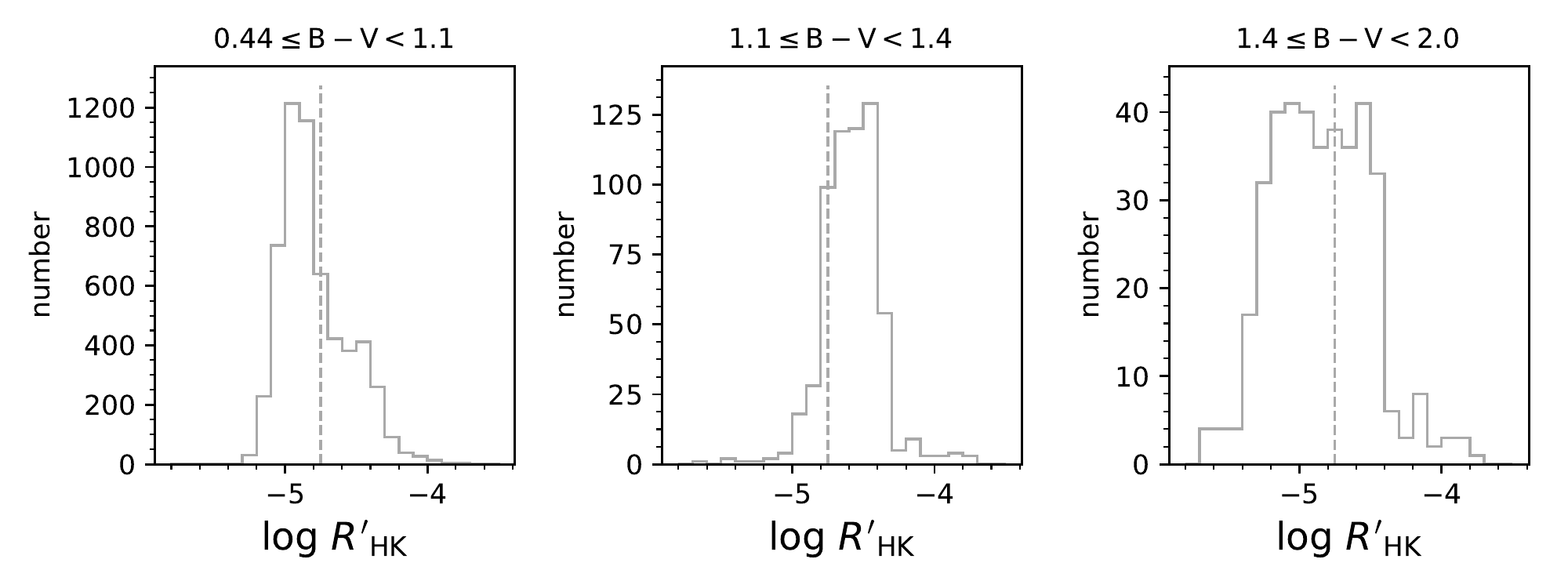}
\caption{
    Normalised distribution of $\log{R'_\mathrm{HK}}$ for different ranges of $B-V$, where the 
  area below each distribution is normalised to 1.
  The vertical dashed lines indicate the approximate position of the Vaughan-Preston gap.
}
\label{hist2}
\end{figure*}
\paragraph{}
For stars with $B-V \leq$0.5, a higher concentration of active stars is seen compared to the stars with $B-V \geq$ 0.5. This effect is due to the rapidly 
increasing rotational velocities of hotter stars, with $B-V \leq$0.5, where the wings around their \ion{Ca}{II} H and K lines can fill in, mimicking an 
active chromosphere \citep{schroeder09}. As a result of this, we see this artificially induced high activity for some stars with $B-V \leq$0.5. As for the 
stars with 0.5$\leq B-V \leq$1.1, the upper and the lower level of activity do not exhibit any significant trend. The Sun also lies in this $B-V$ range, 
and it lies closer to the low chromospheric activity boundary, also known as basal flux \citep{rutten84,schrijver87,rutten91,mittag13,perezmartinez14}. This 
is consistent with previous work by \citet{noyes84}, except for the significant number of stars at intermediate activity level, which has not previously been reported. 
\paragraph{}
For stars with 1.1$\leq B-V \leq$1.4, the basal flux increases linearly with $B-V$, which could indicate a faster spin-down rate. However, the spin-down 
rates for cool stars are not expected to change so dramatically for the spectral types of these stars \citep{reiners12}. It is unclear why these spectral types have 
such a high basal flux level, and further discussion is beyond the scope of this paper. Finally, for cooler stars with $B-V \geq$1.4, the chromospheric 
activity becomes weaker with decreasing $T_\mathrm{eff}$ for later spectral types. This result is consistent with previous work on low-mass stars 
\citep{mohanty03,reiners08}. Although cooler stars are fast rotators, their chromospheric activity is weaker. Such weak chromospheres could be a product of a 
neutral stellar atmosphere that could inhibit magnetic activity \citep{mohanty02,mohanty03}. For completeness, Fig. \ref{rphk_plot} was also plotted for only 
the mean and the median, as shown in Fig. \ref{A3} and Fig. \ref{A4}. The overall shape of the histogram is the same, although we detect a 
slightly higher fraction of very active stars in Fig. \ref{A4}. This is most likely an artificial effect due to flare activity, which influences the mean more 
than the median.
\subsection{Vaughan-Preston gap}
\citet{vaughanpreston80} discovered the lack of intermediately active F and G stars from a sample of 486 stars observed in the Mount Wilson project. We 
investigated the Vaughan-Preston gap based on a sample of 4454 cool stars taken from several surveys. The distribution of activity levels in the right 
panel of Fig.~\ref{rphk_plot} shows some indications of the bimodal distribution of activity proposed by \citet{vaughanpreston80}, although it is less 
distinct than previously found. The region between the two grey lines in Fig. \ref{rphk_plot} indicates the Vaughan-Preston gap reported in \citet{noyes84}. Contrary to
\citet{noyes84}, a considerable fraction of stars are detected in the intermediate-activity region. It was suggested by \citet{vaughanpreston80} that the 
bimodal distribution of activity arises because of two distinct stellar populations: higher values are young and active stars, while lower values are old and 
inactive stars. They also suggested that the gap could either be a representation of different dynamo mechanisms or a statistical bias due to their small 
sample size. According to \citet{durney81}, the gap suggests two different dynamo modes. Alternatively, the gap is explained 
by \citet{noyes84} with three different possibilities: 1) it is not real, 2) there is a true bimodal distribution, but rotation and chromospheric emission 
decrease smoothly with age, or 3) there is a mass-dependent, critical rotation rate where rapid spin-down occurs. They also suggested that the dependence of 
chromospheric activity on either side of the gap on stellar rotation and spectral type questions the idea of two different dynamos. Similar to 
\citet{noyes84}, \citet{rutten87} attributed the gap to a rapid spin-down for stars beyond a certain rotation rate. \citet{rutten87} also remarked that the 
Vaughan-Preston gap is limited to F and G stars and is
absent  in cooler dwarfs. Lastly, the gap might be caused by stars crossing the gap on a very fast 
timescale of $\sim 200 \mbox{Myr}$ \citep{pace09}.
\paragraph{}
In order to investigate this 'bimodality' in more detail, we plot the distribution of stellar activity levels for different ranges of $B-V$ in 
Fig.~\ref{hist2}. For stars with $0.44 \leq B-V \leq 1.1$,  we detect a high concentration of low-activity stars, caused by the decrease in activity as the
star spins down, finally arriving at a minimum activity level. For the F, G, and early-K dwarfs in the range of $0.44 \leq B-V \leq 1.1$, the Vaughan-Preston 
gap is not so apparent. It is not clear if two peaks are noticeable or just one strong peak. Even if the secondary peak does exist, it is not so significant 
as the Vaughan-Preston gap. Moreover, the bimodality might also be fit with a skewed distribution or a Gaussian and skewed distribution.
\paragraph{}
As we move towards late-K and early-M dwarfs in $1.1 \leq B-V \leq 1.4$, the scenario changes quite drastically. We detect a higher concentration of active 
stars on the active side of the Vaughan-Preston gap, and there is a significantly lower number of less active stars. This is the same lack of low-activity K 
dwarfs that was also noted by \citet{mittag13}, which is expected
because \citet{mittag13} shares many stars with our study. One explanation is 
that K dwarfs are in general more active than F and G dwarfs, and their minimum activity level or their basal flux is higher than that of F and G dwarfs.
Whether the lack of inactive K stars arises because they exhibit a higher inactive state or because of other physical origins is unclear.
\paragraph{}
From $1.4 \leq B-V \leq 2.0$, the number of stars that lie on the inactive side of the Vaughan-Preston gap increases, and the 
distribution of activity levels is spread more widely. This is explained by the downtrend of $\log{R'_\mathrm{HK}}$
in this range.
\paragraph{}
For completeness, we show  Fig. \ref{hist2}  with only mean and only median values to determine whether there is some influence from one statistical measurement on 
the other (see Figs. \ref{A3} and \ref{A4} in the appendix). The overall distribution shape of activity levels does not change.
\paragraph{}
Two of the most popular explanations of the Vaughan-Preston gap are stellar spin-down and different dynamo mechanisms. The insignificance of the 
Vaughan-Preston gap suggests that stars spin down from high activity to low activity. Whether there is any rapid spin-down at intermediate activity level is 
not clear. Robust measurements of stellar age and rotation are required to test this theory. To test the idea of two different dynamo mechanisms or geometry, 
 spectropolarimetric observations of these stars and stellar dynamo models need to be combined. The magnetic geometries of these stars are 
currently being investigated using Zeeman Doppler imaging (ZDI) \citep{brown91,donati97}, and the results will be reported in paper II.
\begin{table}
\caption{Common stars for HARPS S-index calibration to the Mount Wilson scale. Column 1 and 2 show the S-index from the Mount Wilson project and HARPS 
spectra, respectively.}
\label{table1}
\centering
\begin{tabular}{c c c}
\hline
Star&S$_\mathrm{MtWilson}$&S$_\mathrm{HARPS}$\\
\hline
\hline
HD10700&0.171& 0.181\\
HD115617&0.164& 0.172\\
HD152391&0.386&0.408\\
HD160346&0.305&0.305\\
HD16160&0.222& 0.233 \\
HD216385&0.142&0.154\\
HD23249&0.138&0.146\\
HD26965&0.208&0.195\\
\hline
\end{tabular}
\end{table} 
\begin{longtab} 
\begin{longtable}{llcccccccccc} 
\caption{{\label{cyclestable}}Cool stars with well determined activity cycles are 
tabulated first, followed by chaotic and multiple cycles, and finally probable cycles. The HD name of the star is shown in column 2, the spectral type 
in  column 3 and the rotation period (days) in column 4. Columns 5, 6, 7 and 8 show the cycle periods (years) and FAPs of the stars 
determined in this work.  The $B-V$ and $\tau_c$ are shown in column 9 and 10. Finally columns 11 and 12 show the Active and Inactive branch classification from \citet{saar99} and the corresponding surveys. MW stands for Mount Wilson and H stands for HARPS.}\\ 
\hline\hline 
no.&name&spectype&P$_\mathrm{rot}$&P$_\mathrm{cyc}$&FAP&P$_\mathrm{cyc}(2)$&FAP(2)&$B-V$&$\tau_c$&branch&survey\\
\hline 
\endfirsthead 
\caption{continued.}\\ 
\hline\hline 
no.&name&spectype&P$_\mathrm{rot}$&P$_\mathrm{cyc}$&FAP&P$_\mathrm{cyc}(2)$&FAP(2)&$B-V$&$\tau_c$&branch&survey\\
\hline 
\endhead 
\hline 
\endfoot 
\multicolumn{12}{c}{\small{CA: Cool stars with clear well defined solar-like activity cycles}}\\ 
\hline
1&Sun$^\dagger$&G2V&25&10.7$\pm$0.08&1.0e-545&..&..&&&..&MW\\
2&HD3651$^\dagger$&K0V&44.0&14.9$\pm$0.2&4.2e-97&..&..&0.85&20.87&I&MW\\
3&HD4628$^\dagger$&K4V&38.5&8.47$\pm$0.05&4.4e-194&..&..&0.89&21.72&I&MW\\
4&HD10476$^\dagger$&K1V&35.2&10.59$\pm$0.1&2.6e-122&..&..&0.836&20.51&I&MW\\
5&HD16160$^\dagger$&K3V&48.0&12.7$\pm$0.11&2.4e-183&..&..&0.918&22.16&I&MW\\
6&HD26965$^\dagger$&K1V&43.0&10.23$\pm$0.07&1.2e-163&..&..&0.82&20.04&I&MW\\
7&HD32147$^\dagger$&K5V&48.0&10.84$\pm$0.15&1.5e-60&..&..&1.049&23.38&I&MW\\
8&HD81809$^\dagger$&G2V&40.2&8.01$\pm$0.05&1.8e-129&..&..&0.655&12.27&I&MW\\
9&HD103095$^\dagger$&K1V&31.0&7.06$\pm$0.05&7.0e-129&..&..&0.754&17.51&I&MW\\
10&HD152391$^\dagger$&G7V&11.43&11.97$\pm$0.1&6.3e-199&..&..&0.749&17.28&A&MW\\
11&HD160346$^\dagger$&K3V&36.4&7.19$\pm$0.04&3.5e-173&..&..&0.594&8.75&I&MW\\
12&HD166620$^\dagger$&K2V&42.4&17.05$\pm$0.2&4.8e-173&..&..&0.876&21.46&I&MW\\
13&HD185144$^\dagger$&K0V&29.0&7.07$\pm$0.14&2.2e-124&..&..&0.786&18.85&I&MW\\
14&HD201091$^\dagger$&K5V&35.7&7.1$\pm$0.05&4.0e-131&..&..&1.069&23.53&I&MW\\
15&HD219834B$^\dagger$&K2V&43.0&9.42$\pm$0.1&2.5e-81&..&..&0.787&18.89&I&MW\\
\hline
\multicolumn{12}{c}{\small{CB: Cool stars with multiple cycles}}\\
\hline
1&HD1835$^\dagger$&G2.5V&7.78&22.0$\pm$0.93&4.2e-44&9.06$\pm$0.15&5.5e-39&0.659&12.51&A&MW\\
2&HD101501&G8V&16.68&23.2$\pm$1.5&3.4e-53&4.5$\pm$0.03&6.4e-39&0.723&16.02&..&MW\\
3&HD149661$^\dagger$&K2V&21.07&15.3$\pm$0.4&8.5e-72&7.7$\pm$0.12&1.2e-42&0.827&20.25&A&MW\\
4&HD190406$^\dagger$&G1V&13.94&2.6$\pm$0.02&1.1e-52&18.7$\pm$0.9&8.8e-50&0.61&9.67&I&MW\\
5&HD20630$^\dagger$&G5V&9.24&22.3$\pm$0.98&6.3e-51&5.7$\pm$0.15&2.9e-11&0.681&13.77&A&MW\\
\hline
\multicolumn{12}{c}{\small{CC: Cool stars with probable activity cycle}}\\
\hline
1&HD10780&K0V&21.7&10.14$\pm$0.31&1.8e-43&..&..&0.804&19.51&..&MW\\
2&HD18256$^\dagger$&FIV-V&3.0&5.89$\pm$0.11&1.2e-27&..&..&0.471&3.11&A&MW\\
3&HD26913$^\dagger$&G8V&7.1&13.3$\pm$0.5&2.8e-41&..&..&0.68&13.71&A&MW\\
4&HD26923&G0V&60.4&8.58$\pm$0.17&1.4e-32&..&..&0.57&7.48&..&MW\\
5&HD37394&K1V&10.74&3.67$\pm$0.07&7.2e-13&..&..&0.84&20.62&..&MW\\
6&HD76151$^\dagger$&G3V&15.0&18.1$\pm$0.76&5.9e-80&..&..&0.661&12.62&I&MW\\
7&HD78366$^\dagger$&G0V&9.67&13.34$\pm$0.38&1.4e-48&..&..&0.585&8.26&A&MW\\
8&HD82443&K0V&5.37&13.84$\pm$3.9&1.8e-21&..&..&0.779&18.58&..&MW\\
9&HD82885$^\dagger$&G8IV-V&18.6&7.69$\pm$0.14&4.0e-31&..&..&0.77&18.212&A&MW\\
10&HD100180&F7V&14.6&3.6$\pm$0.04&6.7e-33&..&..&0.57&7.45&..&MW\\
11&HD115043&G1V&5.86&1.6$\pm$0.02&1.0e-13&..&..&0.603&9.26&..&MW\\
12&HD115383$^\dagger$&F8V&3.33&11.79$\pm$0.28&5.5e-29&..&..&0.585&8.26&A&MW\\
13&HD146233&G2V&22.7&11.36$\pm$1.23&2.1e-02&..&..&0.652&12.10&..&MW\\
14&HD155885$^\dagger$&K1V&21.11&5.83$\pm$0.01&2.2e-33&..&..&0.855&21.0&A&MW\\
15&HD155886$^\dagger$&K1V&20.69&21.6$\pm$1.2&3.23e-23&..&..&0.855&21.0&A&MW\\
16&HD156026$^\dagger$&K5V&21.0&21.3$\pm$0.83&3.4-85&..&..&1.144&24.10&..&MW\\
17&HD165341A$^\dagger$&K0V&19.7&5.07$\pm$0.05&1.3e-48&..&..&0.86&21.11&A&MW\\
18&HD190007&K4V&28.98&5.38$\pm$0.06&9.3e-31&..&..&1.128&23.98&..&MW\\
19&HD201092$^\dagger$&K7V&37.84&11.1$\pm$0.18&3.3e-60&..&..&1.31&25.45&I&MW\\
20&HD20003&G8V&37.1&7.7$\pm$0.5&3.8e-09&..&..&0.77&18.25&..&H\\
21&HD20619&G1.5V&22.3&5.0$\pm$0.2&1.5e-14&..&..&0.65&12.27&..&H\\
22&HD21693&G9IV-V&36.3&7.4$\pm$0.3&6.9e-19&..&..&0.71&15.35&..&H\\
23&HD45184&G2V&21.4&4.9$\pm$0.3&1.8e-23&..&..&0.626&10.59&..&H\\
24&HD7199&K0IV-V&41.0&4.0$\pm$0.3&1.9e-11&..&..&0.55&6.42&..&H\\
25&HD82516&K2V&36.7&4.2$\pm$0.1&1.1e-09&..&..&0.93&22.28&..&H\\
26&HD89454&G5V&21.1&4.0$\pm$0.8&1.2e-07&..&..&0.72&15.71&..&H\\
27&HD157830&G6V&24.3&3.9$\pm$0.5&3.9e-06&..&..&0.68&13.98&..&H\\
28&HD361&G1V&14.1&4.6$\pm$0.2&8.6e-03&..&..&0.624&10.48&..&H\\
29&HD12617&K3V&30.7&4.5$\pm$2.2&4.5e-03&..&..&1.02&23.16&..&H\\
30&HD166724&K0IV-V&30.9&2.4$\pm$0.2&1.8e-04&..&..&1.46&26.69&..&H\\
31&HD21749&K4.5V&34.5&9.1$\pm$6.3&3.4e-10&..&..&1.13&23.99&..&H\\
32&HD154577&K2.5V&45.8&5.9$\pm$0.1&5.4e-66&..&..&0.88&21.71&..&H\\
33&HD88742&G0V&11.3&2.9$\pm$0.1&1.3e-15&..&..&0.59&8.65&..&H\\

\hline 
\end{longtable} 
\tablefoot{The $^\dagger$ symbol shows stars which were previously included in the analysis of \citet{saar99}. The rotation periods for the Mount Wilson stars are taken from \citet{saar99} and \citet{hempelmann16}, except for a few cases as 
follows: HD10780, HD26923, HD3734 \citep{gaidos10}, HD82443 \citep{messina99}, HD100180 \citep{olah16}, HD146233 \citep{petit08}. The rotation periods for the 
HARPS stars are taken from \citet{lovis11}. The $\tau_\mathrm{c}$ used here is calculated based on the empirical relation of \citet{noyes84}, where $\tau_\mathrm{c}$ is a function of $B-V$.\\ 
 
}
\end{longtab}

\section{Chromospheric activity cycles}
It is well known that cool stars other than the Sun also exhibit chromospheric activity cycles \citep{baliunas95}. Since the catalogue consists of multiple 
observations of some stars, it provides a unique opportunity for investigating chromospheric activity cycles. We determined the activity-cycle periods using 
the generalised Lomb Scargle periodogram \citep{zechmeister09}.
\subsection{Cycle period}
Of the 4454 main-sequence dwarfs in the catalogue, only the Mount Wilson and HARPS surveys have multi-epoch observations suitable for a period search (1131 
stars). The majority of these 1131 stars had to be discarded due to the limited number of observations and the non-detection of a periodic signal, resulting in only 
53 stars with identified periods. To investigate stellar activity cycles, we used the generalised Lomb-Scargle periodogram \citep{zechmeister09}. Unlike the 
standard Lomb-Scargle method \citep{lomb76,scargle82}, this technique accounts for the observational uncertainties and allows a floating mean. This makes 
the period determination more robust and less prone to aliasing (for a detailed analytic solution to the generalised Lomb-Scargle technique, we refer to 
\cite{zechmeister09}). The strongest peak in the periodogram is considered as the period. The uncertainties in the cycle period are calculated by propagating 
the errors in the periodogram equation.
\begin{figure}
\centering
\includegraphics[scale=0.4]{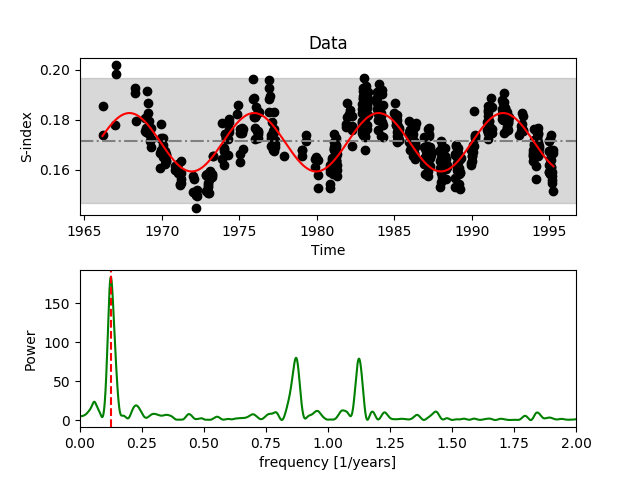}
\caption{Example of a cycle period categorised as CA (HD81809). The top plot shows the data in black, and the red curve shows the fit to the strongest peak in the periodogram shown in the bottom plot. The shaded grey area shows the span of the solar chromospheric activity from one 
minimum to the maximum shifted to the centre at the star's mean value. The dashed horizontal line shows the mean of the data points. The strongest 
frequency is marked by the dashed red line in the bottom plot.}
\label{CA_periodogram}
\end{figure}
\begin{figure}
\centering
\includegraphics[scale=0.4]{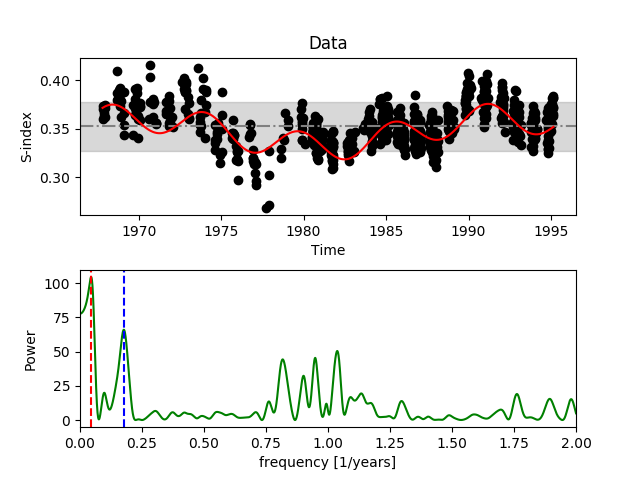}
\caption{Example of a cycle period categorised as CB (HD20630). The top plot shows the data in black circles. The red curve is the fit to the two strongest 
peaks in the periodogram shown in the bottom plot. The shaded grey area shows the span of the solar chromospheric 
activity from minimum to the maximum, shifted to the centre at the star's mean value. The dashed horizontal line shows the mean of the data points. The two strongest peaks are marked by the 
dashed red and blue lines in the bottom plot.}
\label{CB_periodogram}
\end{figure}
\begin{figure}
\centering
\includegraphics[scale=0.4]{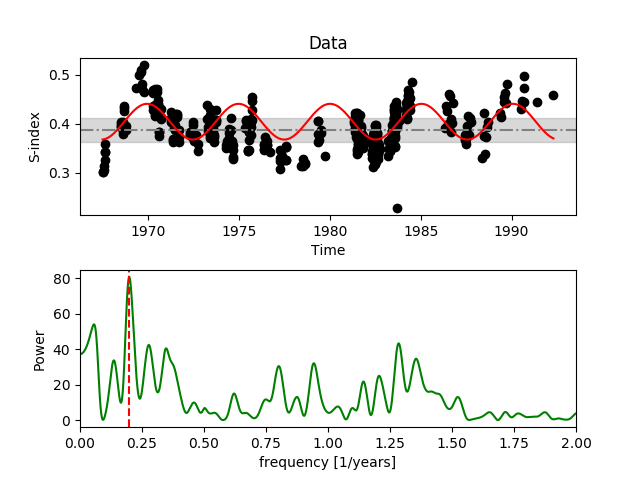}
\caption{Example of a cycle period categorised as CC (HD165341A). The top plot shows the data in black, and the red curve shows the fit to the strongest peak in the periodogram shown in the bottom plot. The shaded grey area shows the span of the solar chromospheric activity from one 
minimum to the maximum, shifted to the centre at the star's mean value. The dashed horizontal line shows the mean of the data points. The strongest peak is marked by the dashed red line in the bottom 
plot.}
\label{CC_periodogram}
\end{figure}
\begin{figure*}
\centering
\sidecaption
\includegraphics[scale=0.55]{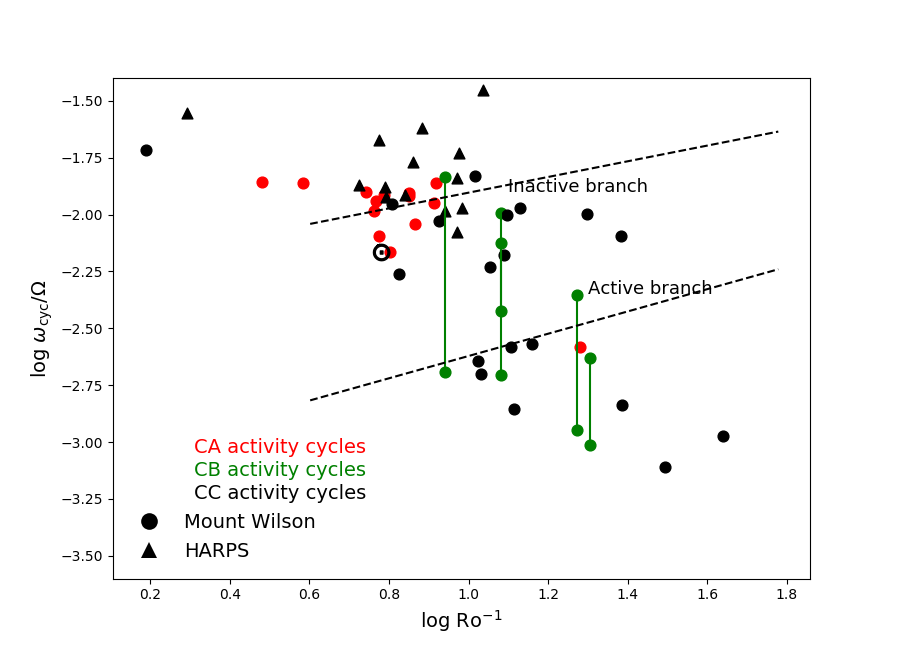} 
\caption{ $\omega_\mathrm{cyc}/\Omega$ vs. $Ro^{-1}$ for the group of stars in Table \ref{cyclestable}, shown in log scale. The red symbols are activity cycles classified as CA, green symbols are activity cycles
classified as CB, and black symbols are activity cycles classified as CC. The circles denote Mount Wilson stars, and triangles represent HARPS stars.
For multiple cycles, the second cycle periods (Pcyc2) are connected to the filled symbols by a horizontal line. The Sun is shown as $\sun$. The active and inactive 
branches from \citet{saar99} are shown as black dotted lines
(see Section 4 for details on the selection criteria).}
\label{saar_all}
\end{figure*}
\paragraph{}
To determine the statistical significance of the period, we also calculated the false-alarm probability \textup{(\textup{FAP})} of the detected signal. For our purposes,
we also performed a  `visual selection' where we inspected each time series and only included stars with observations spanning at least four years. Any 
cycle length longer than 25 years was excluded because that is the limit of a full cycle recovery based on the observational time span of the surveys. We also 
excluded cycle periods that were shorter than two years because with increasingly shorter period, it becomes difficult to distinguish between the rotation period and the activity-cycle period. Additionally, the sampling is not robust enough to confirm periods shorter than two years. 
There is one exception to this rule, however: HD115043. This star has a probable cycle period of 1.6 years and a low \textit{\textit{\textup{FAP}}}. This is the 
only star in the sample to show a cycle period shorter than two years.
\subsection{Activity-cycle classification}
Determining the presence of an activity cycle based on the period search algorithm alone comes with caveats. Since the only robust point of reference of an 
activity cycle is the solar cycle, we assumed the stellar activity cycle to be sinusoidal. However, in the case of the Sun, several other 
cycles are also detected, although they are much weaker than the dominant 11-year cycle. Multiple cycle periods instead of one clear activity cycle have
also been detected in a 
few other cool stars \citep{baliunas95,vitense07,saar99,olah16}. {\citet{olah16} found that multiple cycles are most likely to be detected in fast-rotating 
stars.}
\paragraph{}
Our analysis showed that while some stars exhibit solar-like activity cycles, some show irregular cycles; irregular cycles were also reported by 
\cite{baliunas95}. We divided the stars into three categories: Class A (CA), Class B (CB), and Class C (CC); they are listed in Table \ref{cyclestable}. Four factors 
were considered for classifying the stars: (i) visual confirmation of the signal, where the periodic signal can be easily detected; (ii) amplitude of the 
periodic signal, where the amplitude is at least similar to the solar case; (iii) repetition of the signal over the time series, to confirm that it is 
indeed a regular cycle; and (iv) the $\text{FAP}$ of the periodic signal. A considerable difference in the $\text{FAP}$ is seen for stars belonging to the three groups. A brief
description of the three categories is given below. 
\paragraph{\textbf{CA: Stars with well-defined cycles}} 
Stars that show a clear solar-like sinusoidal activity cycle are included in this category. {Figure \ref{CA_periodogram} shows an example where the cycle 
period is repeated in the long-term observations and the amplitude of the cycle is at least similar to the solar case. This provides additional confidence 
in the activity-cycle period.} As shown in Table \ref{cyclestable}, the majority of these stars (a total of 15 stars including the Sun) are slowly rotating 
stars with rotation periods equivalent to or longer than the solar rotation period. Only HD152391 has a rotation period shorter than 12 days, and is not 
classified as an inactive branch star by \citet{saar99} either. The time series and corresponding fit from the period search for these well-defined cycles are 
shown in Fig. \ref{A}. The stars belonging to this category were
taken from the Mount Wilson survey. A secondary period close to one year is also detected in a 
majority of these stars. This is not a cycle period but a harmonics seen due to the nature of the seasonal observations.
\paragraph{\textbf{CB: Stars with multiple or chaotic cycles}}
This group of stars shows chaotic or multiple cycles. {An example of such a star is shown in Fig. \ref{CB_periodogram}. There are 
only five stars in this group, and they are taken from the NSO Mount Wilson survey. Although a few other stars exhibited possible multiple periodicity, their 
$\text{FAP}$s were not reliable for the secondary periods. Additionally, they could be aliases and were discarded. {A further detailed analysis is required for some 
individual stars to confirm their multiple periods.} The time series and the corresponding periodograms are shown in Fig \ref{B}.} 
\paragraph{\textbf{CC: Stars with unconfirmed cycles}}
The stars in this category exhibit possible solar-like cycle periods. {Thirty-three stars fall in this category}. The confidence level in the $\text{FAP}$, the amplitude of the 
signal, or the sampling are too low for them to be included in the CA category. An example of such a cycle is shown in Fig. \ref{CC_periodogram}. Further 
observations are required to confirm the nature of these activity cycles. Both Mount Wilson and HARPS stars can be found in this category. Figs. \ref{C} and 
\ref{D} show the time series of these stars. 
\section{Long-term evolution of chromospheric activity versus rotation}
Previously, \citet{saar99,vitense07} have shown that there is a possible relationship between activity-cycle period and rotation period, and the stellar 
activity cycles classify stars into two branches. We investigate the activity cycle versus rotation relation for the cycle periods determined in this work.
\subsection{$\omega_{cyc}/\Omega$ versus $Ro^{-1}$}
To investigate the rotation versus activity-cycle period relation, we list our stars in Table \ref{cyclestable} in the $\omega_{cyc}/\Omega$ versus Ro$^{-1}$ plane 
(see Fig. \ref{saar_all}), similar to Fig. 1 of \citet{saar99}. Following the definition of \citet{noyes84}, the $Ro$ number used here is determined using an 
empirically determined 
$\tau_{c}$. The $\tau_\mathrm{c}$ used by \citet{saar99} in their analysis is theoretically determined. Since the empirical $\tau_\mathrm{c}$ of 
\citet{noyes84} and theoretical $\tau_\mathrm{c}$ of \citet{saar99} are in good agreement (see Fig 3. in \citet{saar99}), we used the \citet{noyes84} 
definition. Based on the Mount Wilson long-term survey, previous work investigating activity cycles in cool stars has shown that stars are segregated into 
two branches, the active and inactive branch \citep{saar99,brandenburg98,vitense07}. They also noted that there is a clear lack of stars in between these two branches. 
\paragraph{}
It is worth noting that the following stars from \citet{saar99} are not included in our analysis: HD98230B is excluded because
of the lack of available data, HD187691
is excluded because of the lack of significant variability in its S-index values, and HD154417 was excluded as the measured cycle periods fall outside of our 
selection boundary. The stars marked with a $\text{dagger}$  in Table \ref{cyclestable} are the stars that have previously
been published by \citet{saar99}. \footnote{Not all 
the entries in Table 1 of \citet{saar99} were included in their analysis. They stated that they only included stars with weights greater than 0.} The cycle 
periods of some of our common stars vary from the periods in Table 1 of \citet{saar99}. This discrepancy is most likely caused by the difference in the data 
used because \citet{saar99} also used results from \citet{radick98}, which are not included in our analysis. Additionally, we used the new release of the Mount 
Wilson sample, which has more observations. Overall, our results are consistent with those of \citet{baliunas95}, except in a few cases. 
In this version of the generalised Lomb-Scargle periodogram, measurement errors are weighted and an offset is introduced to account for the floating mean. These modifications make it a more robust algorithm than the periodogram \citep{scargle82} used in 
\citet{baliunas95}.
\paragraph{}
Figure \ref{saar_all} shows that stars with clear activity cycles
(CA), including our Sun, all lie on the inactive branch, except for one star that lies on the
active branch of \citet{saar99}. Almost all stars with cycles classified as CA have $\log \mathrm{Ro}\leq$ 1. Mount Wilson stars whose cycles fall in the 
CB and CC category mostly exhibit $\log\mathrm{Ro}\geq$ 1, and some of them populate the intermediate region between the active and inactive branch.  Two of the five CB 
stars exhibit cycle periods that alternate between the inactive branch and the active branch. The primary and secondary periods of HD149661 
lie in the intermediate region. The primary and secondary cycles of HD1835 and HD20630 lie outside the active branch. The CC stars
are spread throughout the rotation and cycle period plane. The HARPS CC stars lie close to the inactive branch. The Mount Wilson CC stars, on the other hand, can be found 
around the inactive branch, the active branch, and the intermediate region.
\paragraph{}
When the three classifications (CA, CB, and CC) are combined, no clear distinction between the inactive and active branch is detected. Figure \ref{saar_all} 
also shows that log $\omega_{cyc}/\Omega$ decreases with increasing log $Ro^{-1}$. This linear trend is not a physical effect and is caused by the selection 
bias. Based on our selection criteria, the stars with cycle periods shorter than 1.6 years and greater than 25 years were not included in our analysis. This 
creates the effect in Fig \ref{saar_all}.  Our results show that stellar activity cycles could lie in the intermediate region between the active and inactive
branches. It is hard to discern where the inactive branch ends and the active branch begins. Figure \ref{saar_all} does not show any clear indication of 
the positive slopes that have been proposed for the activity branches, unless we only focus on the stars classified as CA. If we do not include the CB and CC stars, then 
only the inactive branch stars are left. This strengthens our argument that the classification into active or inactive branches
might not be a true representation of 
the complex relationship between cycle periods and stellar rotation.
\paragraph{}
Our results are in qualitative agreement with the recent results of \citet{olspert17}, where the authors find an indication of an inactive branch, but not 
an active branch by re-analysing the Mount Wilson sample. Furthermore, recent numerical studies of global convective dynamo simulations for a wide range of 
rotation rates find no indication of activity branches \citep{viviani17, warnecke17}. In these studies, the distribution of dynamo cycles also agrees 
well with the one found in this work, see in particular Figure 9 of \citet{warnecke17}.
\begin{figure*}
\centering
\sidecaption
\includegraphics[scale=0.5]{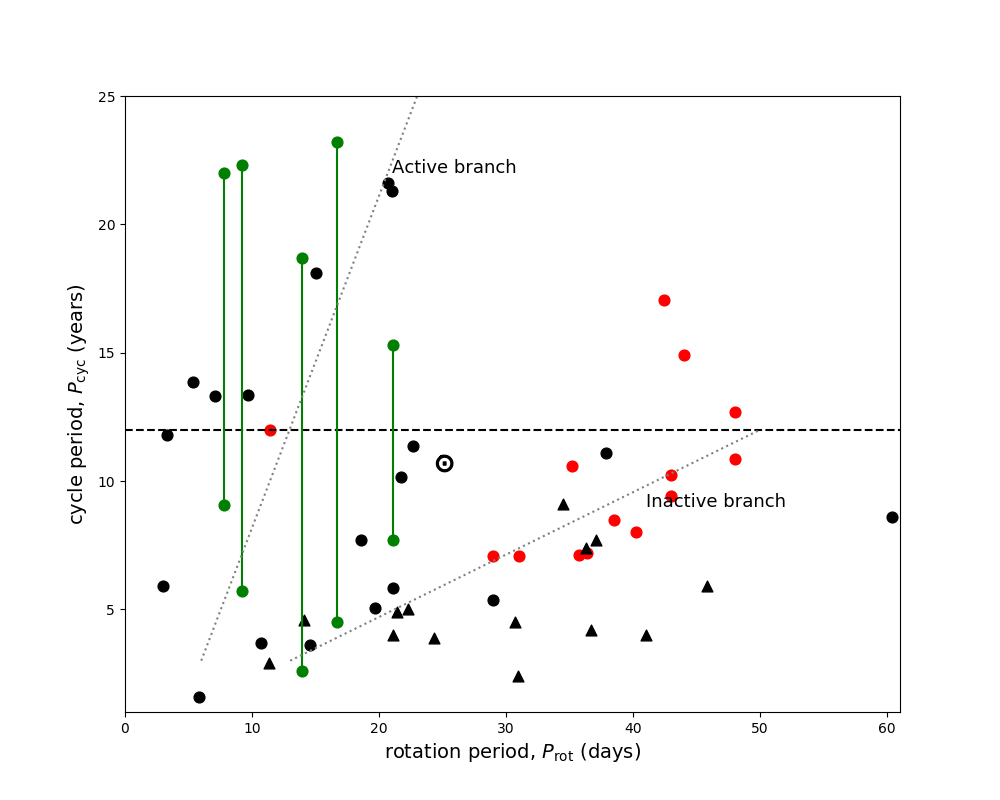}
\caption{Activity-cycle period in years as a function of rotation period in days for stars in Table \ref{cyclestable}. The symbols are as same as Fig \ref{saar_all}. The black dotted lines show the active and inactive branch according to
\citet{vitense07}. The black horizontal line marks the midpoint of the maximum cycle length of 25 years.}
\label{rotcyc}
\end{figure*}
\subsection{$P_\mathrm{cyc}$ versus $P_\mathrm{rot}$}
In order to investigate the magnetic activity cycles versus rotation relation independent of any empirical trend, we plot the activity-cycle period as a function of 
rotation period similar to Fig. 1 in \citet{vitense07}. Figure \ref{rotcyc} shows that the well-defined cycles, classified as CA, mostly lie on the inactive 
branch of stellar cycles. However, the active branch of stellar cycles is less distinct than other studies. Additionally, for the first time, the region 
around the Sun is populated. We also detect stellar cycles that lie in the lower regions of the inactive branch. However, these stars are classified as CC, 
and care should be taken as further observations are required to be certain.
\paragraph{}
Previous work by \citet{vitense07} showed that stellar activity cycles are clearly separated into active and inactive branches, and activity cycles
of stars on the active branch can migrate to the inactive branch. Our results show that a limited number of stars exhibit multiple cycles, and only 
two of these show cycle periods migrate from the inactive branch to the active branch. The multiple periods of one star lie in the intermediate region, where the 
Sun is accompanied by a few other stars.  Furthermore, the inclusion of the HARPS stars shows that stars could also lie in the region below the inactive 
branch, which makes it hard to clearly classify activity cycles into branches. Figure \ref{rotcyc} also shows that stellar activity cycles cannot be clearly 
divided into two branches. Only stars classified as CA exhibit a possible linear trend with rotation. The linear trend disappears when CB and CC stars are 
also included in the analysis. Finally, if we only consider the CA stars in Fig. \ref{rotcyc}, then the linear trend on the inactive branch could have a 
different slope (see Appendix \ref{CAfig}). Under these conditions, the Sun appears to lie much closer to the inactive branch. In both cases, whether we 
consider the three different classifications or the CA classification
alone, the active branch appears to be very weak and non-existent.
\subsection{Multiple periodicity in some stars}
In addition to the clear solar-like cycles detected in the CA stars, multiple cycles were also observed in Figs. \ref{saar_all} and \ref{rotcyc}. Multiple cycles 
are observed on stars rotating faster than the Sun. The solar cycle used here is determined from chromospheric activity measurements. Long-term monitoring 
of solar activity has shown that although the solar cycle is cyclic in nature, it is not purely periodic. The Sun also exhibits several cycles that are not 
as dominant as the 11-year cycle \citep[see][for more details on solar activity]{solanki06}. For example, the long-term modulation of the sunspot numbers also 
indicates an 80-year cycle known as the Gleissberg cycle, as well as periodicities of 51.34 years, 8.83 years, and 3.77 years that are seen in the difference 
in activity levels between the two hemispheres \citep[see][]{2016AJ....151...70D}. When viewed as a star from the ecliptic, the north-south asymmetries cancel,
and we are left with only the 11-year activity cycle (and periodicities associated with its longer term modulation). However, when viewed from outside of the 
ecliptic, the north-south asymmetries no longer exactly compensate for each other, and a number of additional frequencies are expected. The amplitude of the Sun's 
activity  cycle, as seen in sunspots, is considerably higher than the amplitude of these other cycles, but this comparison is based on data that 
include the current grand maximum \citep{2004Natur.431.1084S}, and it does not take into account that the amplitude of the other cycles varies strongly in time 
\citep{2016AJ....151...70D}. In other stars, and at other times for the Sun, it is possible that the relative amplitudes of all these cycles could be 
different. Hence, the presence of multiple cycles in some stars is not surprising and is most likely caused by various factors: inclination, complex 
magnetic field  geometry, observational epoch, or any combination of the three.
\paragraph{}
We ran the generalised Lomb Scargle periodogram on the solar S-index data but were only able to detect the activity cycle of 10.7 years. This cycle period is in agreement with the 
sunspot cycle of 11 years. We did not detect any multiple periodicity on the S-index time series. The data were not sampled at a high enough cadence to 
detect these shorter and much weaker additional periodicities.
\section{Comparison with results from Zeeman Doppler imaging}
One theory explaining the division of activity cycles into two branches is that two different magnetic geometries are sustained by two different dynamo mechanisms 
\citep{vitense07}. In order to investigate the magnetic geometry of stars in Figs. \ref{saar_all} and \ref{rotcyc}, the Zeeman Doppler imaging (ZDI) technique 
has to be used. ZDI is the only technique that can reconstruct the large-scale surface magnetic geometry of cool stars. One way of comparing if stars have 
similar magnetic geometry is to investigate if their surface magnetic field geometry is poloidal or toroidal. ZDI reconstructs magnetic field geometries in 
the spherical harmonics frame, which allows us to express the magnetic field in its poloidal and toroidal field components.
\paragraph{}
The magnetic field geometry of stars along the inactive and active branches was studied by \citet{see16} on 12 ZDI targets, only
5 of which had 
observations spanning multiple epochs. \citet{see16} showed that the stars that lie on the active branch exhibit strong toroidal geometry. However 
strong poloidal geometry is also detected for a few active branch stars \citep{hackman16,rosen16}. For active stars with multi-epoch observations, 
the toroidal fraction exhibits strong modulation over time and sometimes changes to strong poloidal field components, although one active branch star, HD78366, shows 
a strong poloidal field that remains consistently poloidal over several years of observations \citep{morgenthaler11}, (Jeffers et al. 2018 (in prep)).
\paragraph{}
Stars on the inactive branch, on the other hand, exhibit strong poloidal geometry. However, the sample size of stars on the inactive branch is much smaller: there are only four stars. Additionally, it is not fully known if there is any modulation in the poloidal fraction over time. Only one star was observed over multiple 
epochs; this was 61 Cyg A \citep{borosaikia16}. This star shows a slight increase in its toroidal fraction over its activity cycle. Young fast-rotating 
solar-type stars such as HD206860 \citep{borosaikia15} and $\kappa$ Ceti \citep{nascimento16}, however, exhibit a complex large-scale magnetic geometry with a significant
toroidal component. For comparison, the Sun has a strong poloidal field, and its large-scale field is more than 90$\%$ poloidal at the surface 
\citep{vidotto16}.
\paragraph{}
Recent work by \citet{metcalfe16} investigated the magnetic topology of a handful of solar analogues that have also been observed using the \textit{Kepler} 
satellite. The results of \citet{metcalfe16} show that there is a difference in field geometry for stars above and below the Vaughan-Preston gap and also for 
stars that lie on the active and inactive branches. However, one should be careful while interpreting the results, as the Vaughan-Preston gap and the 
active and inactive branch do not refer to the same phenomena, as shown by Fig. \ref{simple}, where the gaps in the two cases are not the same. It
is clearly shown that stars that lie on the inactive branch (red symbols) do not necessarily lie on the inactive side of the gap, which would be the case if the 
Vaughan-Preston gap and the inactive and active branch were the same. The Vaughan-Preston gap is produced by the lack of stars with intermediate 
chromospheric activity, whereas the separation into inactive and active branches is made as a result of the lack of stars with activity-cycle periods that lie in the empty region 
near the Sun. Furthermore, the magnetic geometry sample used in \citet{metcalfe16} is very limited. The stars they included only show limited epochs 
and  in some cases only one observation epoch. This might be highly misleading in this context, as the poloidal and toroidal fraction for a 
single star is known to change quite significantly over time, and the sample is too small to have any statistical significance. 
\paragraph{}
The magnetic field topology changes even for
the Sun from a more simple dipolar field to a more complex higher-order field during one activity cycle. 
For solar analogues the field geometry could change dramatically from one month to the next \citep{jeffers17}. {For example, the field geometry could 
exhibit a strong poloidal field in one epoch, but turn into a toroidal field in a matter of a few months \citep{see15,jeffers17}.  Hence, it is very 
important to monitor stars over the long term to understand whether the magnetic geometry is indeed different for stars with different levels of activity. 
Understanding the evolution of magnetic field geometry during the dynamo-operated activity cycle of a star is necessary to be able to determine whether the dynamo 
is indeed different for stars in the cycle period-rotation plane.}
\section{Conclusions}
We introduced a chromospheric activity catalogue of 4454 cool stars. We explored the Vaughan-Preston gap for the stars in our catalogue. 
We also investigated the relationship between activity-cycle period and rotation period  in a selected sample of stars. Our main conclusions are listed below.
\begin{itemize}
\item
The ratio of chromospheric to bolometric flux, $\log R'_\mathrm{HK}$, as a function of $B-V$ shows that the Vaughan-Preston gap is not as significant as 
previously thought; this distinction was made possible by our very large catalogue of stars. If the gap is indeed not real, then it suggests that main-sequence stars begin their 
lifetimes with strong activity, and as they spin down, they become less active, finally settling down with a quiescent activity level without any 
sudden break at intermediate activity. Robust measurements of stellar age and rotation are required for a 
reliable conclusion. Finally, ZDI  magnetic geometry reconstructions of stars across the Vaughan-Preston gap combined with stellar dynamo models are required 
to investigate if active and inactive stars indeed have two different dynamo mechanisms. 
\item 
Stars with well-defined clear activity cycles, cycles classified as CA, predominantly lie on the inactive branch, and they exhibit a possible linear relation 
with rotation. Stars with cycles classified as CB (multiple cycles) and CC (probable cycles) are scattered on the activity period-rotation plane, even filling
the gap near the Sun, making the active branch less distinct. When all classifications of stars are considered, no linear relationship is detected between cycle 
period and rotation. The weakening of the active branch classification is still prevalent when the CC stars are not included. Our conclusions still stand when 
we include only the CA stars. Finally, the Sun's position in the new cycle period-rotation plot indicates that it is not an exception and that its dynamo is most 
likely not in a transitional phase, as previously suggested.
\end{itemize}
\begin{acknowledgement}
We thank  A. A. Vidotto, M. J. K\"apyl\"a, D. Schleicher, M. Zechmeister, and V. See for the fruitful discussions that immensely improved the paper. S.B.S, C.J.M, and S.V.J acknowledge support from the SFB 963 (Projects A4 and A16) funded by the DFG. 
S.B.S also acknowledges funding via the Austrian Space Application Programme (ASAP) of the Austrian Research Promotion Agency (FFG) within ASAP11, the 
FWF NFN project S11601-N16 and the sub-project S11604-N16project.
J.W. acknowledges funding from the People Programme (Marie Curie Actions) of the European Union's Seventh Framework Programme (FP7/2007-2013) under REA grant agreement No.\ 623609. We also thank the Mount Wilson NSO project: the {$\mathrm{HK\_Project\_v1995\_NSO}$} data derive from the Mount Wilson Observatory HK Project, which 
was supported by both public and private funds through the Carnegie Observatories, the Mount Wilson Institute, and the Harvard-Smithsonian Center for 
Astrophysics starting in 1966 and continuing for over 36 years.  These data are the result of the dedicated work of O. Wilson, A. Vaughan, G. Preston, 
D. Duncan, S. Baliunas, R. Radick and many others.This research has made use of the VizieR catalogue access tool, CDS, Strasbourg, France. The original 
description of the VizieR service was published in A$\&$AS 143, 23.
\end{acknowledgement}

\bibliographystyle{aa}
\bibliography{ref}
\begin{appendix}
\begin{figure*}
\section{S-index versus $B-V$}
\includegraphics{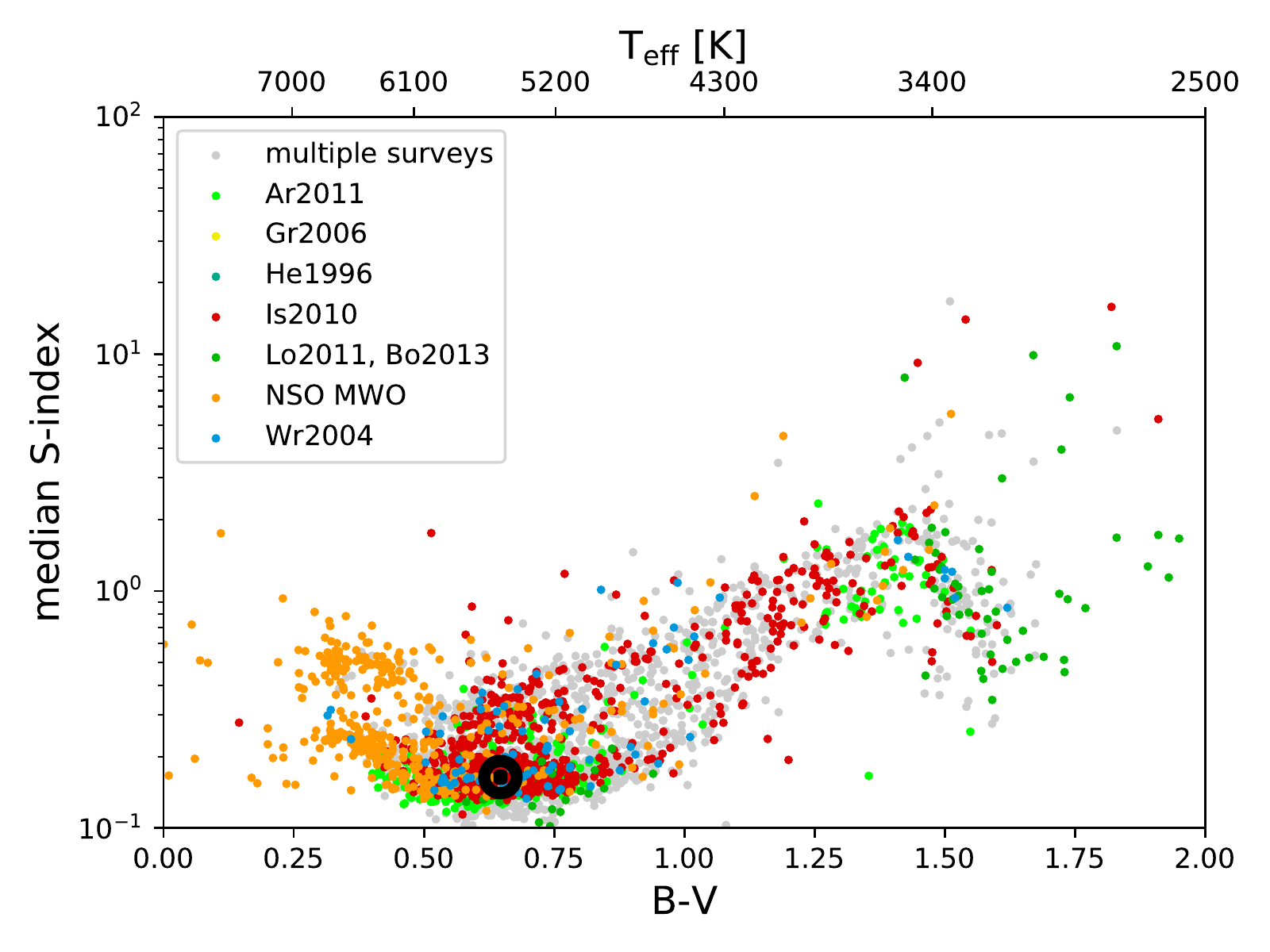}\\
 \caption{Median $S_\mathrm{MWO}$ versus $B-V.$
   The symbols and colour palette are the same as in Fig \ref{S}.
   The Sun at minimum activity is shown by the black $\sun$ symbol.
 }
 \label{A1}
\end{figure*}
\begin{figure*}
\includegraphics{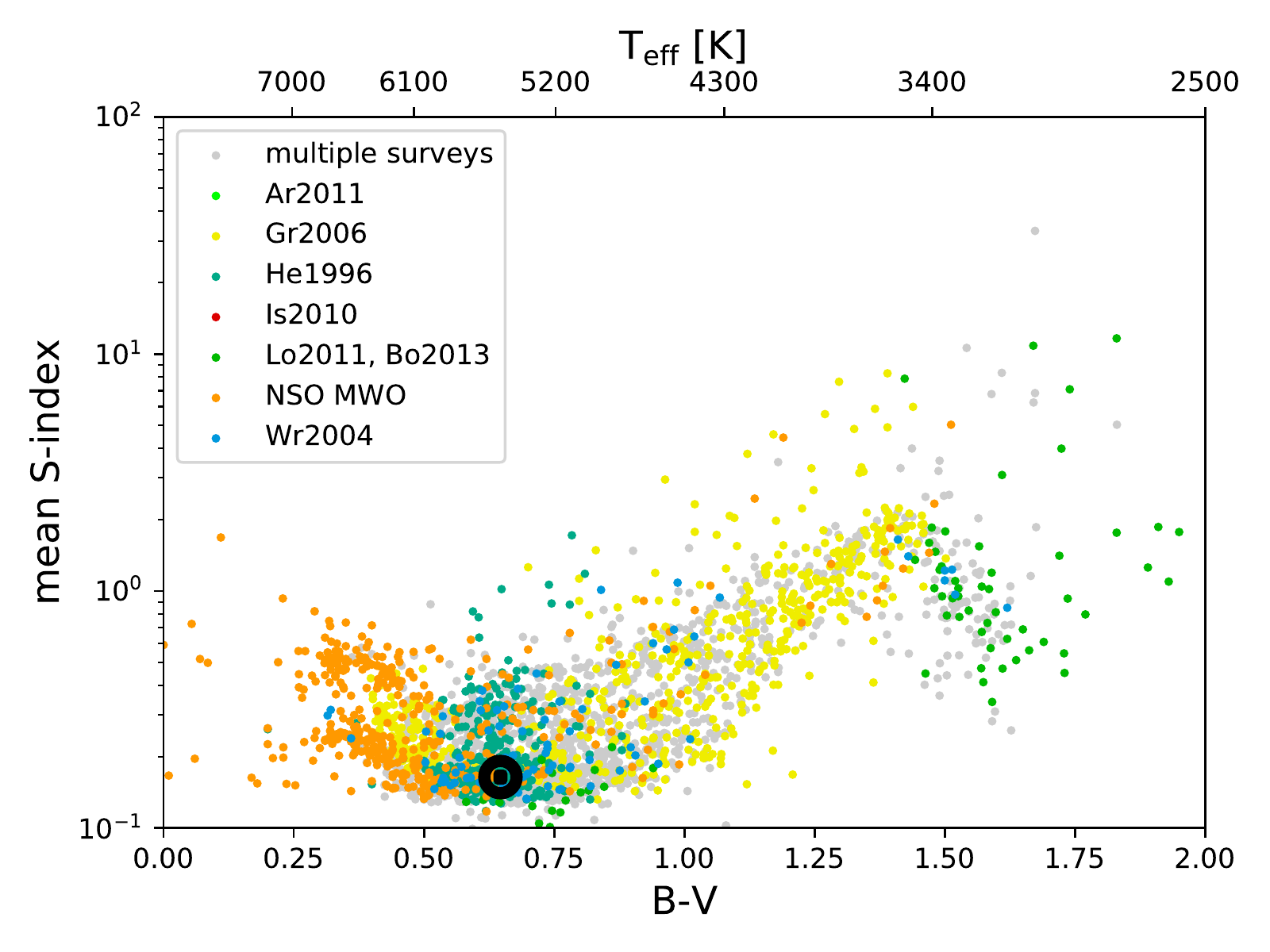}\\
 \caption{Mean $S_\mathrm{MWO}$ versus $B-V.$
   The symbols and colour palette are the same as in Fig \ref{S}.
   The Sun at minimum activity is shown by the black $\sun$ symbol.
 }
 \label{A2}
\end{figure*}
\begin{figure*}
\section{$\log R'_\mathrm{HK}$ versus $B-V$}
\centering
\includegraphics{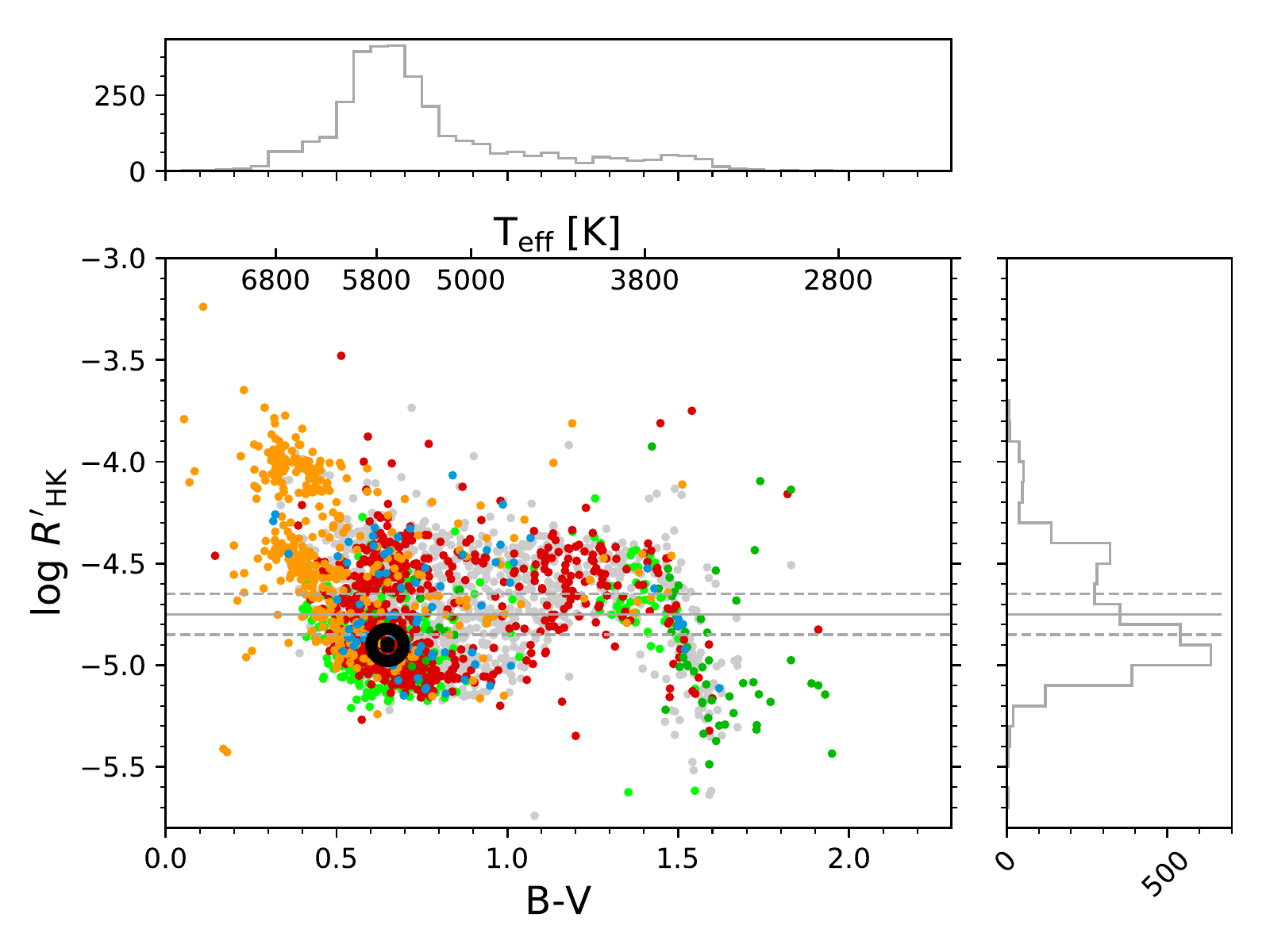}\\
\includegraphics[scale=0.7]{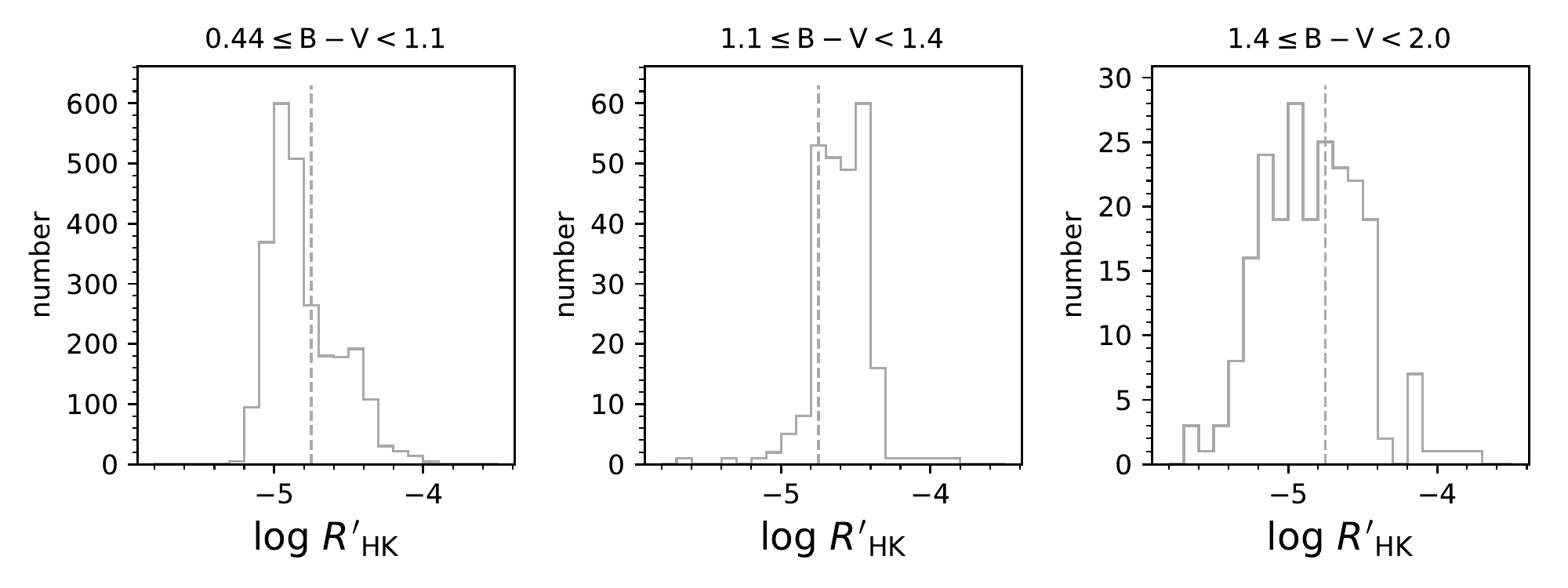}
 \caption{
  \textit{Top}: Ratio of median chromospheric \ion{Ca}{II} H and K flux
   to bolometric flux, $\log R'_\mathrm{HK}$,
   vs. $B-V$.
   The symbols and colour palette are the same as in Fig \ref{S}.
   The Sun at minimum activity is shown by the black $\sun$ symbol.
   The dashed grey lines indicate the Vaughan-Preston gap, as seen in Fig. 2
   of \citet{noyes84}. The histogram on the right shows the distribution of the median chromospheric 
   activity, and the histogram at the top shows the distribution of $B-V$ for stars in the 
   catalogue. \textit{Bottom}: Normalised distribution of the
median $\log{R'_\mathrm{HK}}$ for different 
   ranges of $B-V$, where the area below each distribution is normalised to 1. 
   The vertical dashed lines indicate the approximate position of the Vaughan-Preston gap.
 }
\label{A3}
\end{figure*}

\begin{figure*}
\centering
\includegraphics{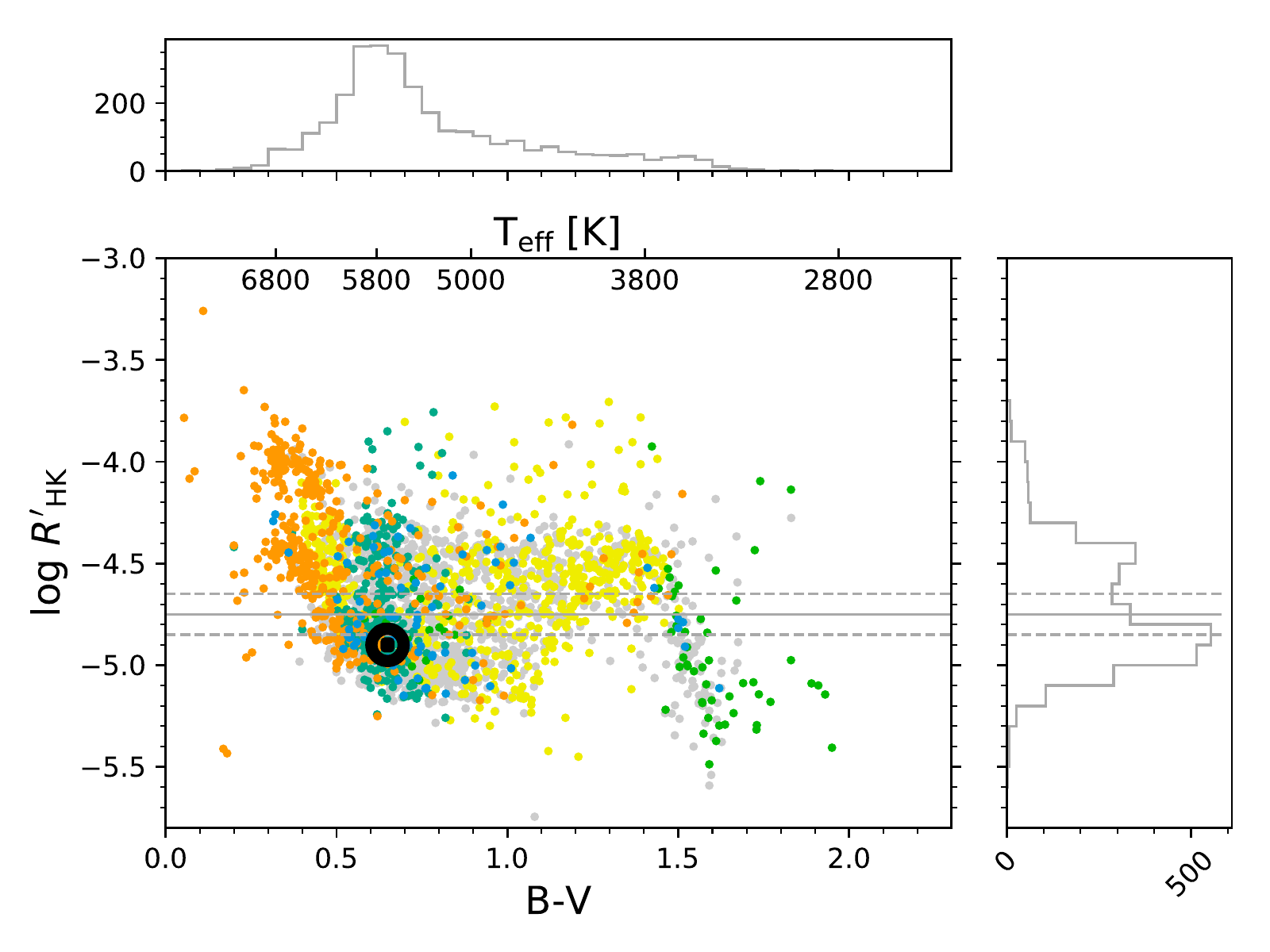}\\
\includegraphics[scale=0.7]{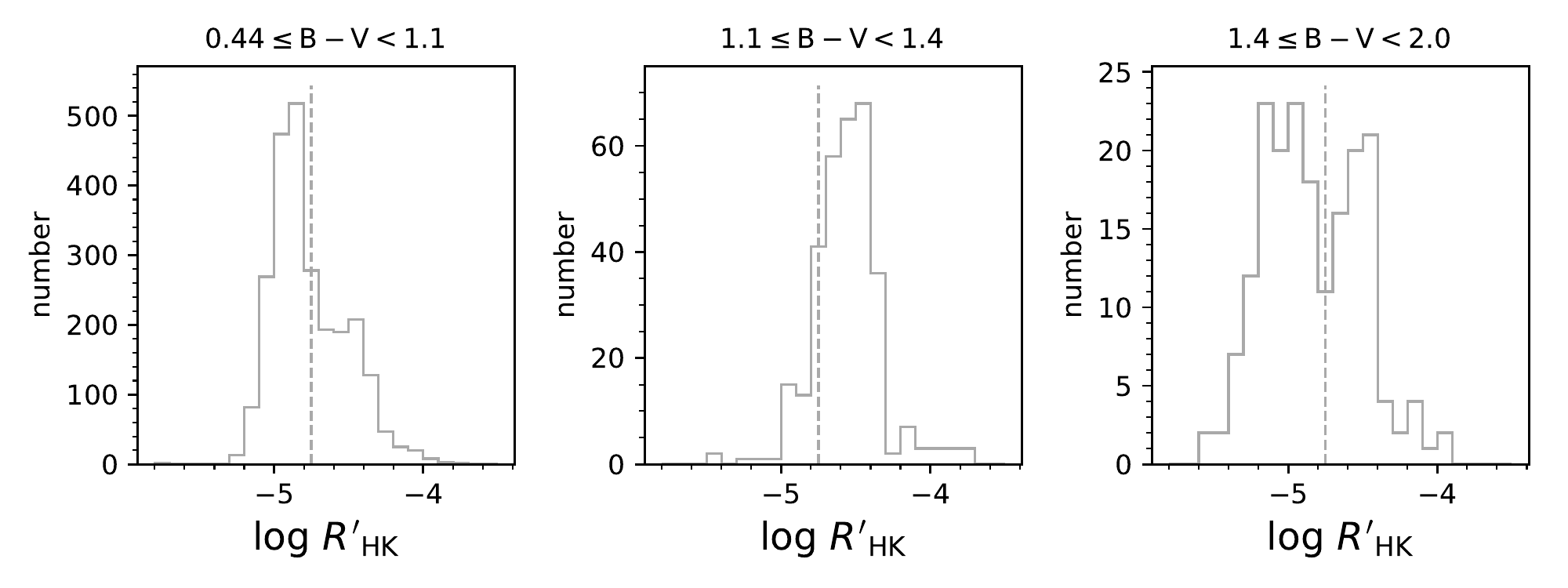}
 \caption{
  \textit{Top}: Ratio of mean chromospheric \ion{Ca}{II} H and K flux
   to bolometric flux, $\log R'_\mathrm{HK}$,
   vs. $B-V$.
   The symbols and colour palette are the same as in Fig \ref{S}.
   The Sun at minimum activity is shown by the black $\sun$ symbol.
   The dashed grey lines indicate the Vaughan-Preston gap, as seen in Fig. 2
   of \citet{noyes84}. The histogram on the right shows the distribution of the median chromospheric 
   activity, and the histogram at the top shows the distribution of $B-V$ for stars in the 
   catalogue. \textit{Bottom}: Normalised distribution of the
mean $\log{R'_\mathrm{HK}}$ for different 
   ranges of $B-V$, where the area below each distribution is normalised to 1. 
   The vertical dashed lines indicate the approximate position of the Vaughan-Preston gap.
 }
\label{A4}
\end{figure*}

\begin{figure*}
\section{Chromospheric activity cycles and the respective periodograms for CA, CB, and CC stars}
\centering
\subfloat[SUN]{\includegraphics[scale=0.32]{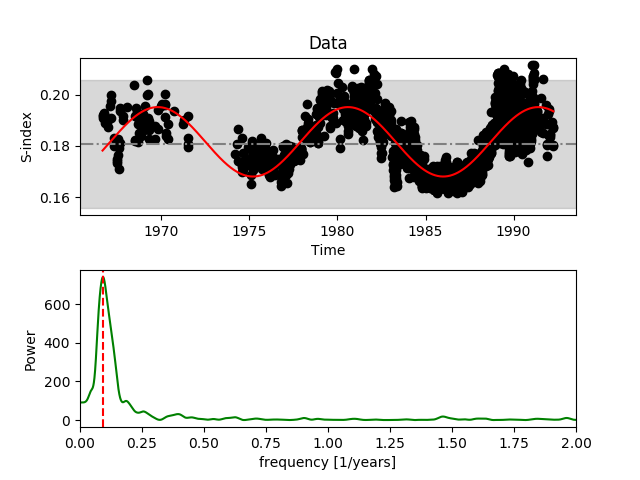}}
\subfloat[HD3651]{\includegraphics[scale=0.32]{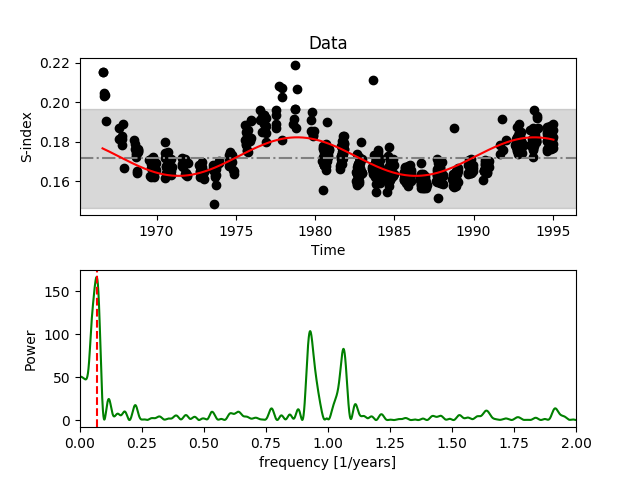}}
\subfloat[HD4628]{\includegraphics[scale=0.32]{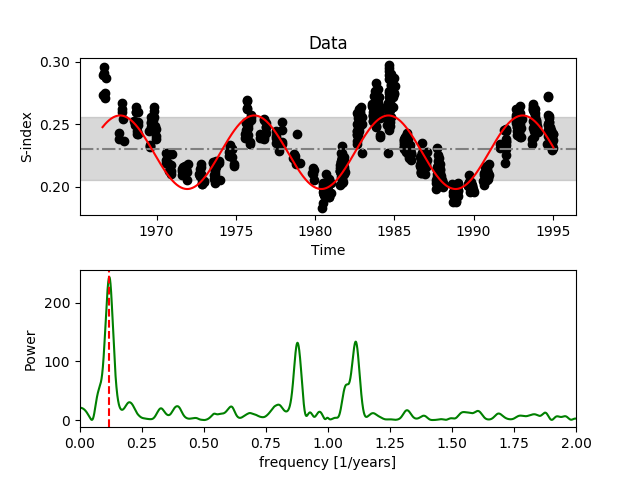}}\\
\subfloat[HD10476]{\includegraphics[scale=0.32]{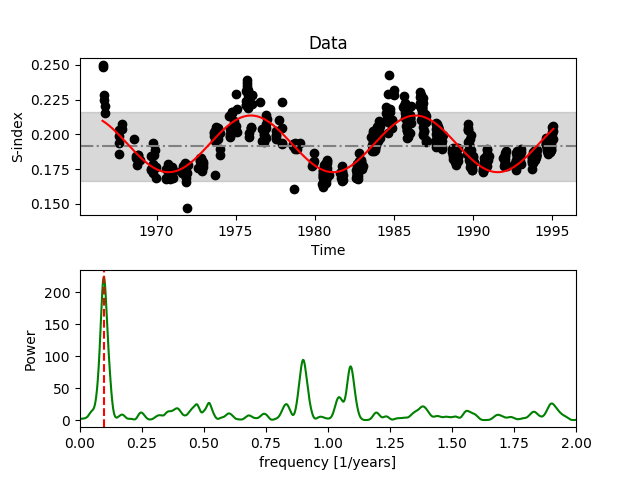}}
\subfloat[HD16160]{\includegraphics[scale=0.32]{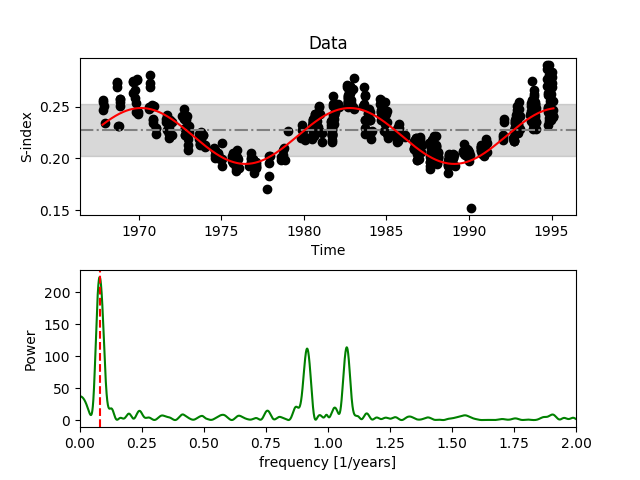}}
\subfloat[HD26965]{\includegraphics[scale=0.32]{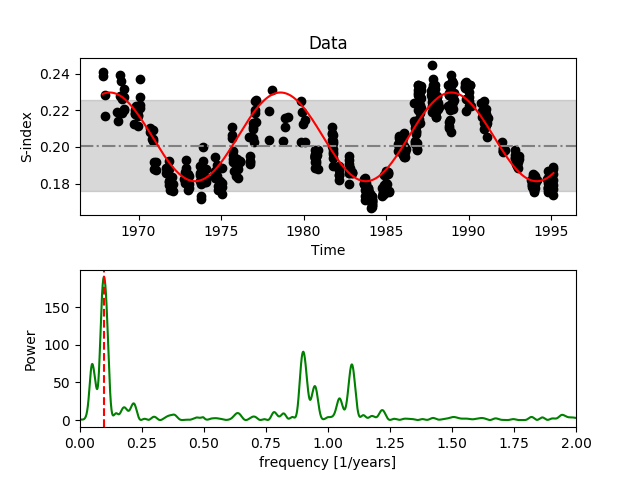}}\\
\subfloat[HD32147]{\includegraphics[scale=0.32]{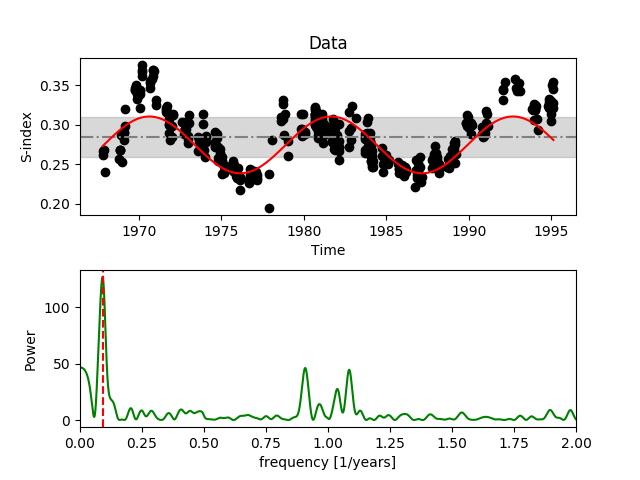}}
\subfloat[HD81809]{\includegraphics[scale=0.32]{HD81809.png}}
\subfloat[HD103095]{\includegraphics[scale=0.32]{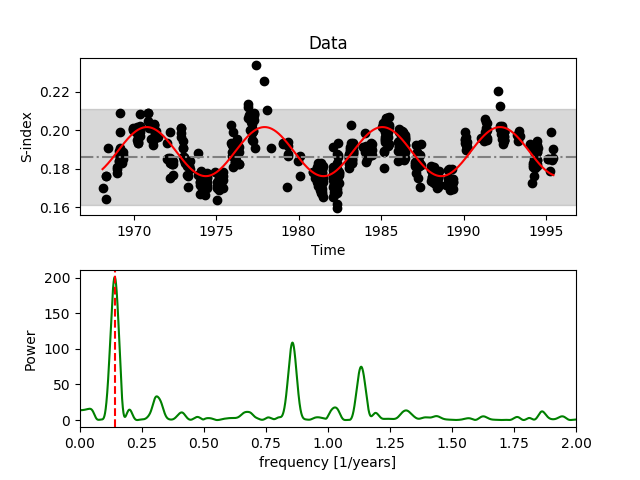}}\\
\subfloat[HD152391]{\includegraphics[scale=0.32]{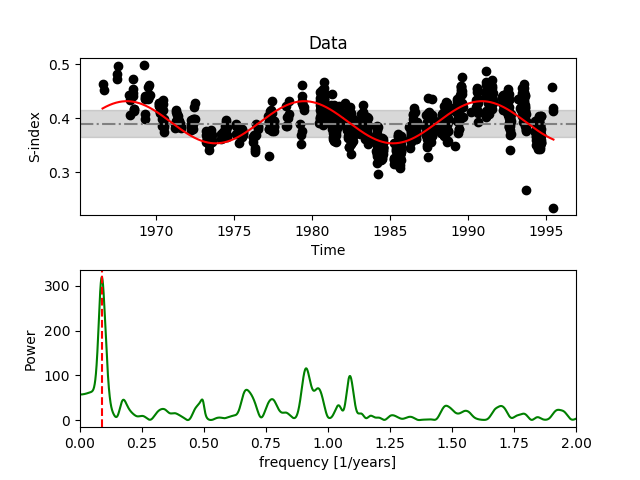}}
\subfloat[HD160346]{\includegraphics[scale=0.32]{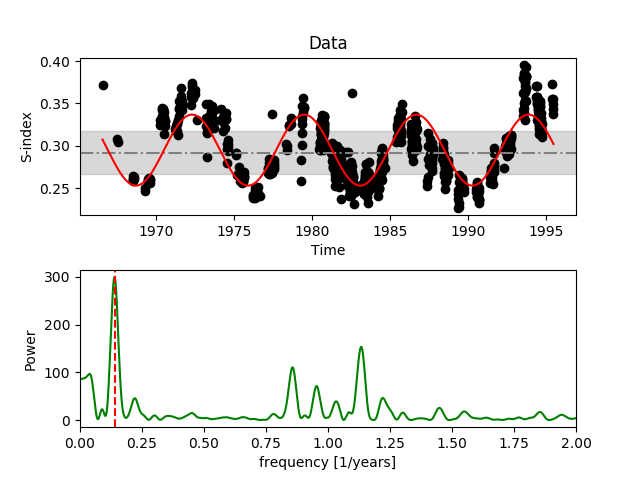}}
\subfloat[HD166620]{\includegraphics[scale=0.32]{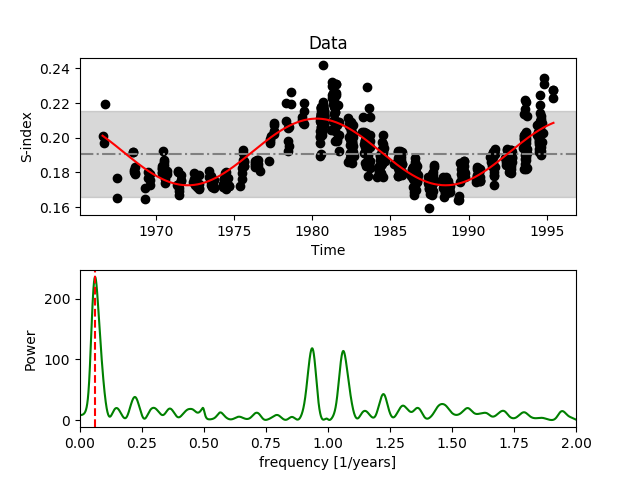}}\\
\subfloat[HD185144]{\includegraphics[scale=0.32]{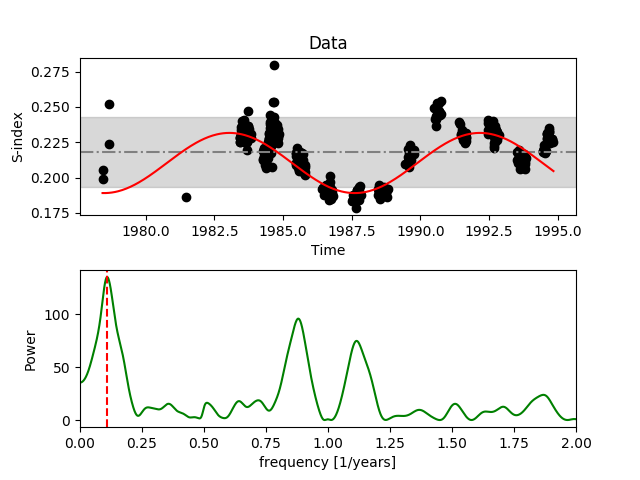}}%
\subfloat[HD201091]{\includegraphics[scale=0.32]{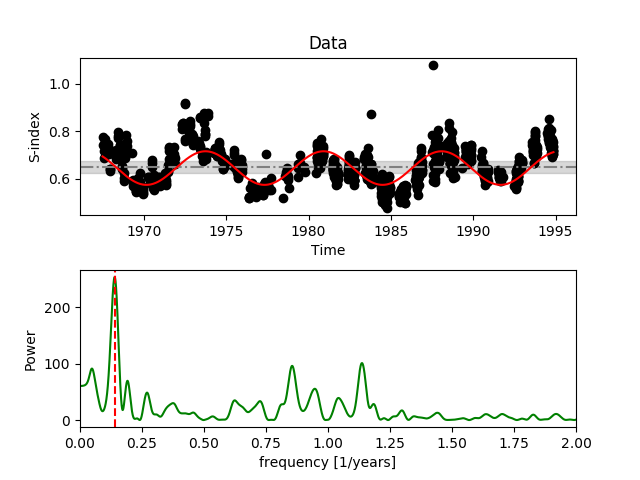}}
\subfloat[HD219834B]{\includegraphics[scale=0.32]{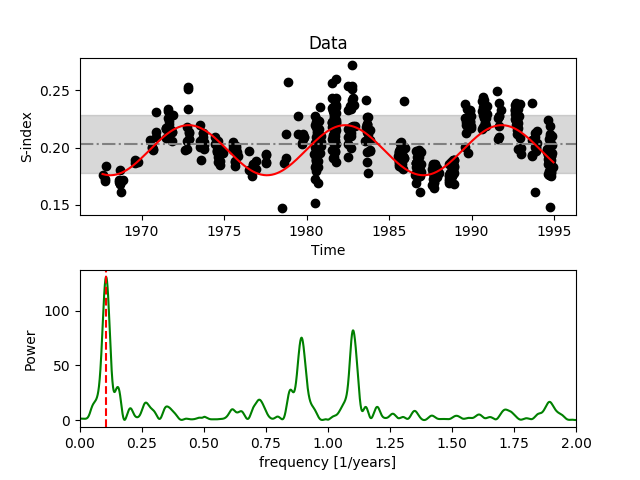}}
\caption{Cool stars (Mount Wilson) with clear solar-like cycles.}
\label{A}
\end{figure*}

\begin{figure*}
\centering
\subfloat[HD1835]{\includegraphics[scale=0.32]{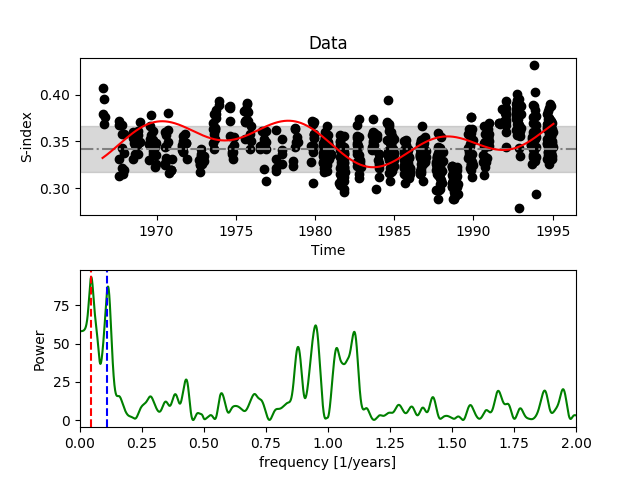}}
\subfloat[HD101501]{\includegraphics[scale=0.32]{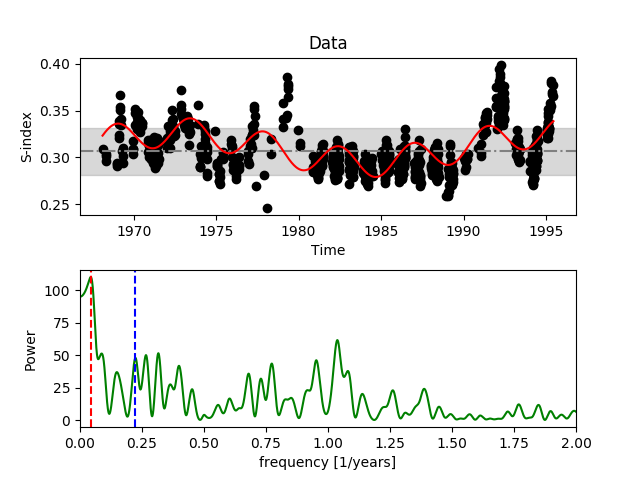}}
\subfloat[HD149661]{\includegraphics[scale=0.32]{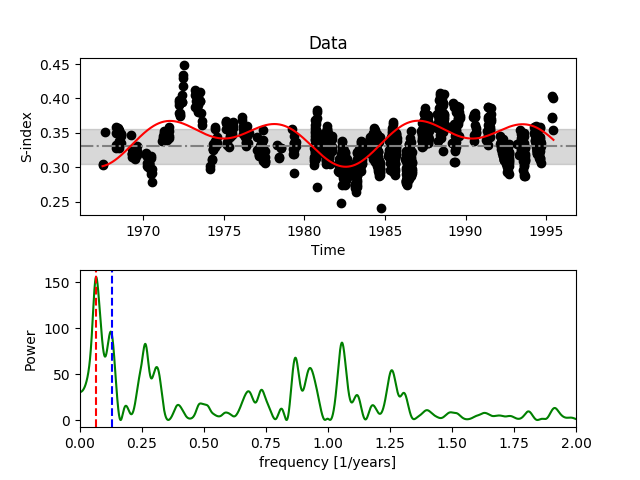}}\\
\subfloat[HD190406]{\includegraphics[scale=0.32]{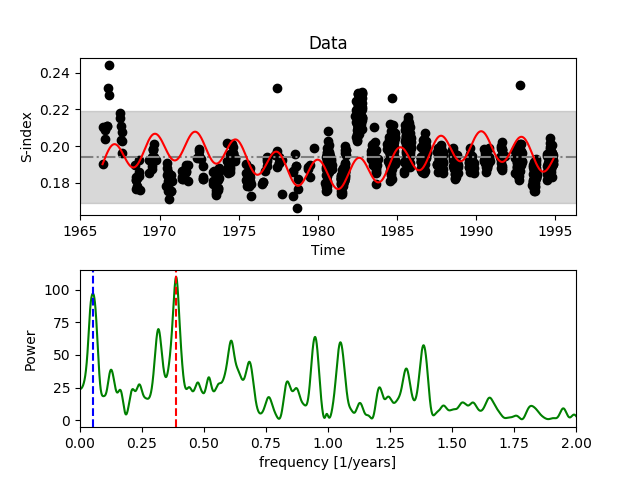}}
\subfloat[HD20630]{\includegraphics[scale=0.32]{HD20630_v2.png}}
\caption{Cool stars (Mount Wilson) with multiple cycles. }
\label{B}
\end{figure*}

\begin{figure*}
\centering
\subfloat[HD10780]{\includegraphics[scale=0.32]{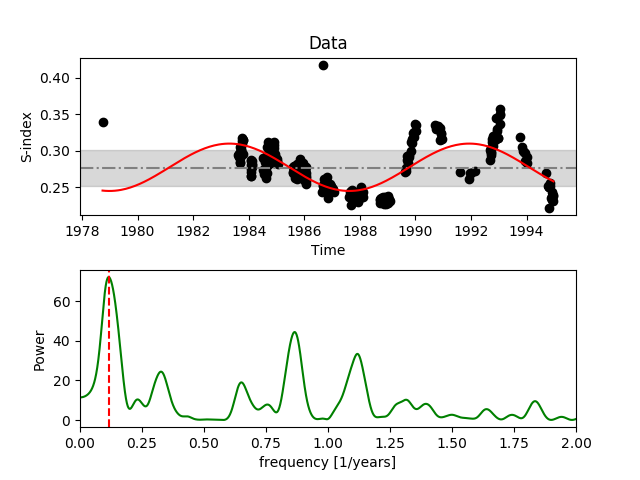}}
\subfloat[HD18256]{\includegraphics[scale=0.32]{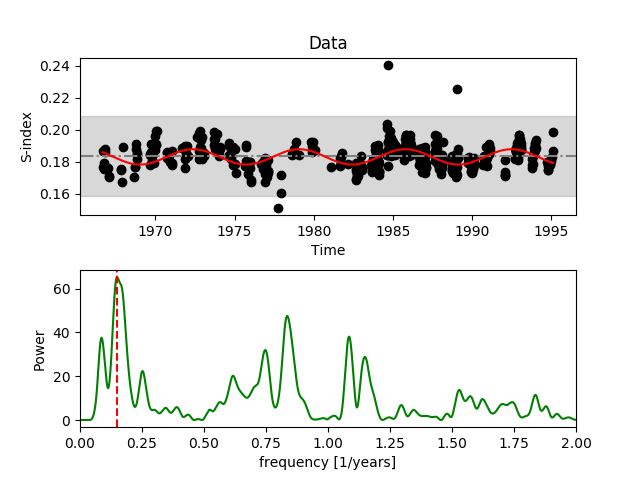}}
\subfloat[HD26913]{\includegraphics[scale=0.32]{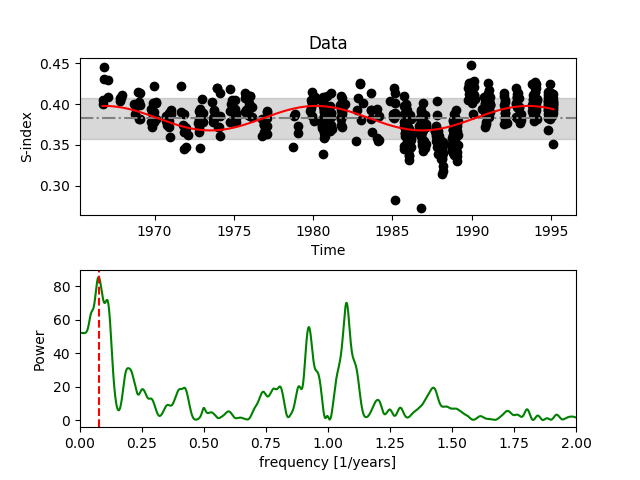}}\\
\subfloat[HD26923]{\includegraphics[scale=0.32]{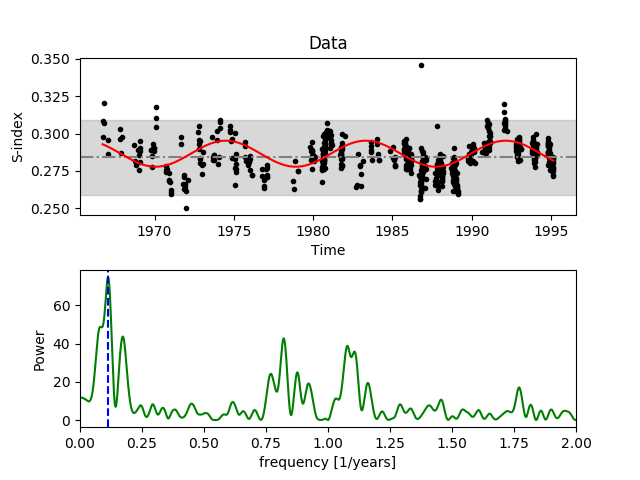}}
\subfloat[HD37394]{\includegraphics[scale=0.32]{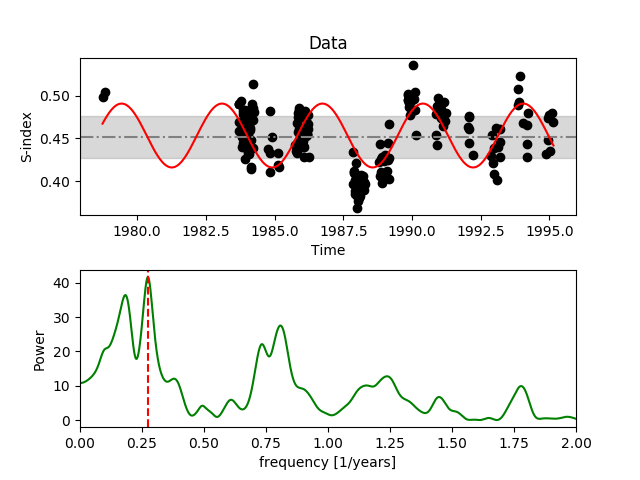}}
\subfloat[HD76151]{\includegraphics[scale=0.32]{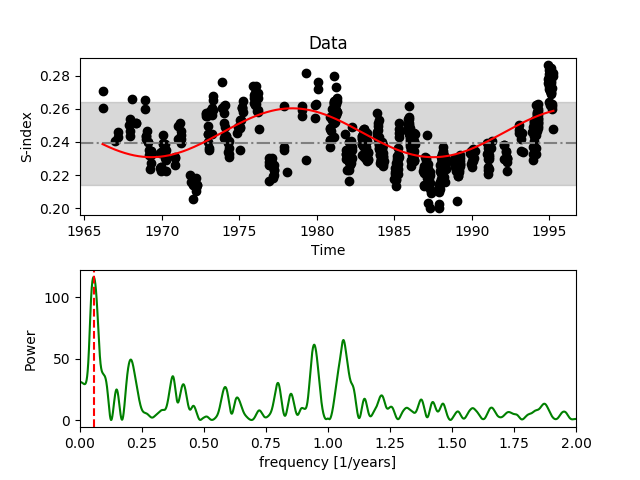}}\\
\subfloat[HD78366]{\includegraphics[scale=0.32]{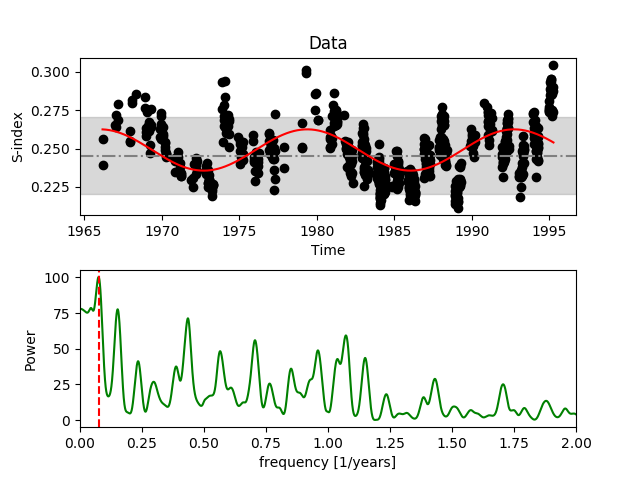}}
\subfloat[HD82443]{\includegraphics[scale=0.32]{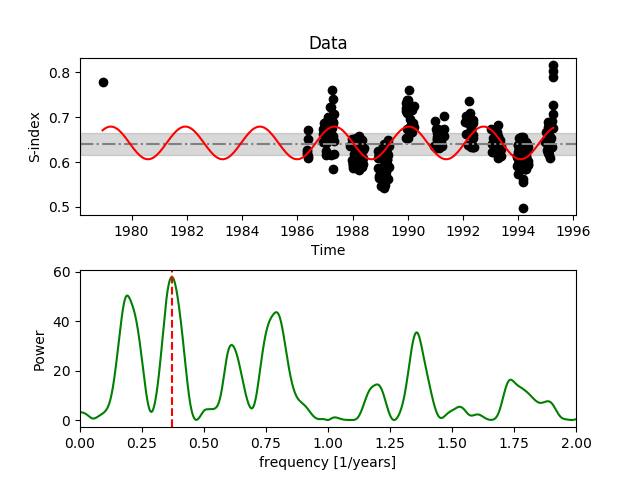}}
\subfloat[HD82885]{\includegraphics[scale=0.32]{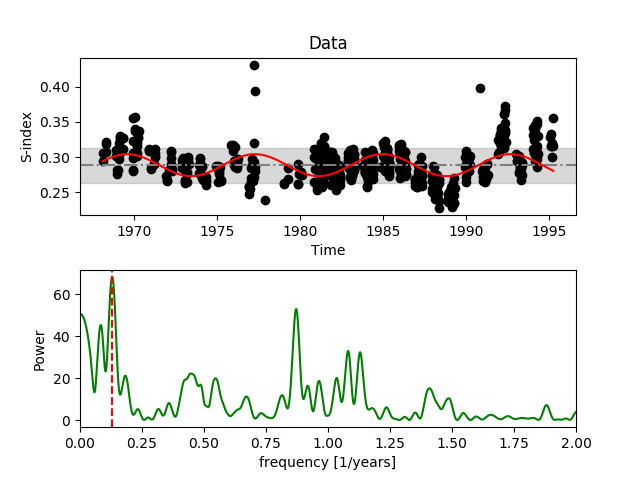}}\\
\subfloat[HD100180]{\includegraphics[scale=0.32]{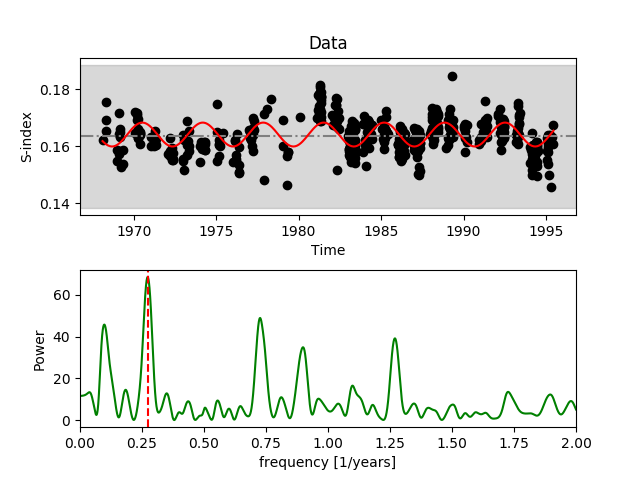}}
\subfloat[HD115043]{\includegraphics[scale=0.32]{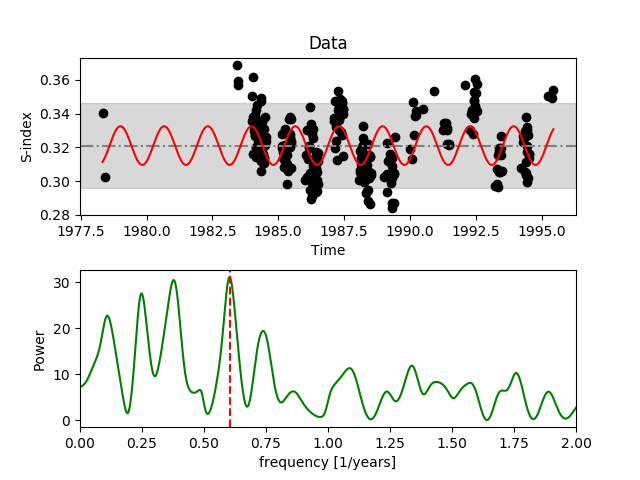}}
\subfloat[HD115383]{\includegraphics[scale=0.32]{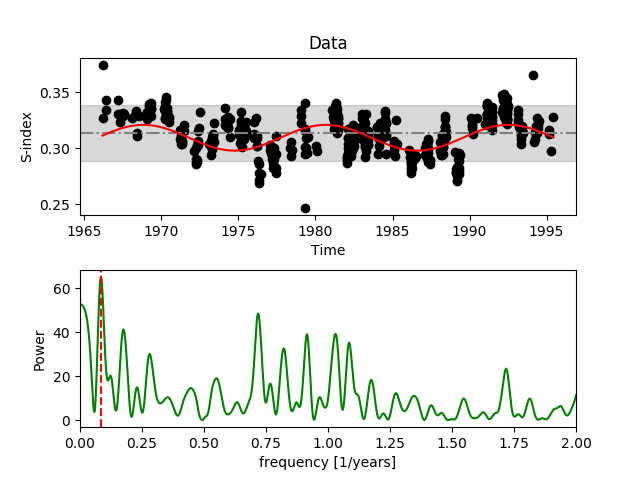}}\\
\subfloat[HD146233]{\includegraphics[scale=0.32]{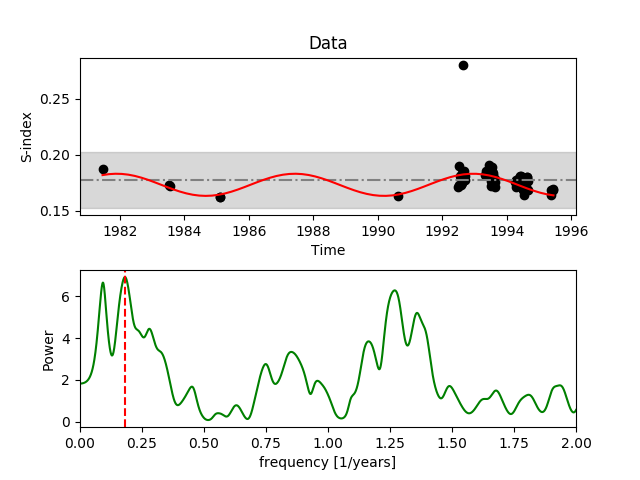}}
\subfloat[HD155885]{\includegraphics[scale=0.32]{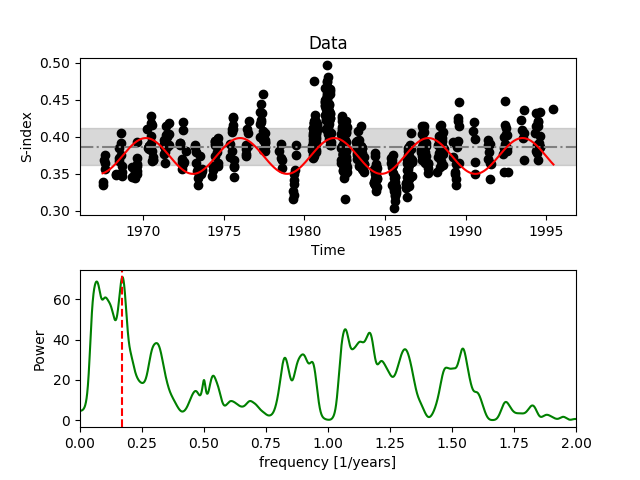}}
\subfloat[HD155886]{\includegraphics[scale=0.32]{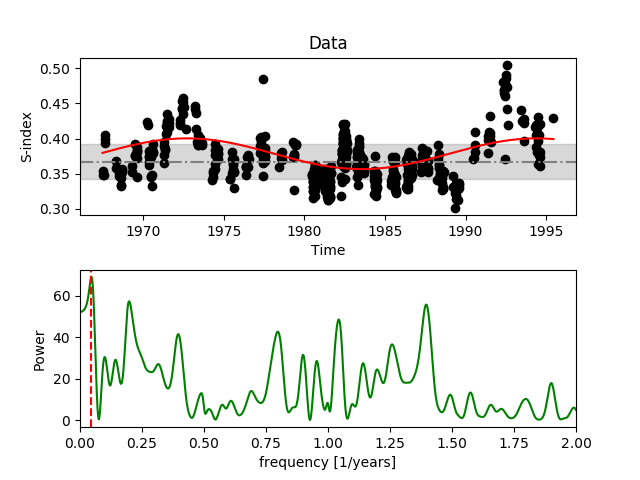}}
\end{figure*}
\begin{figure*}
\centering
\subfloat[HD156026]{\includegraphics[scale=0.32]{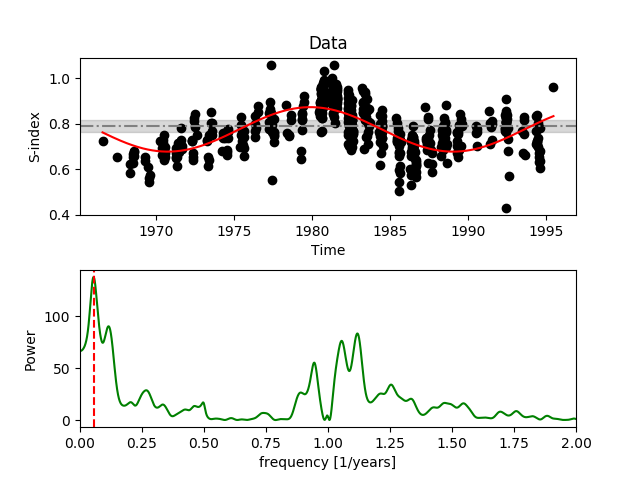}}
\subfloat[HD165341A]{\includegraphics[scale=0.32]{HD165341A.png}}
\subfloat[HD190007]{\includegraphics[scale=0.32]{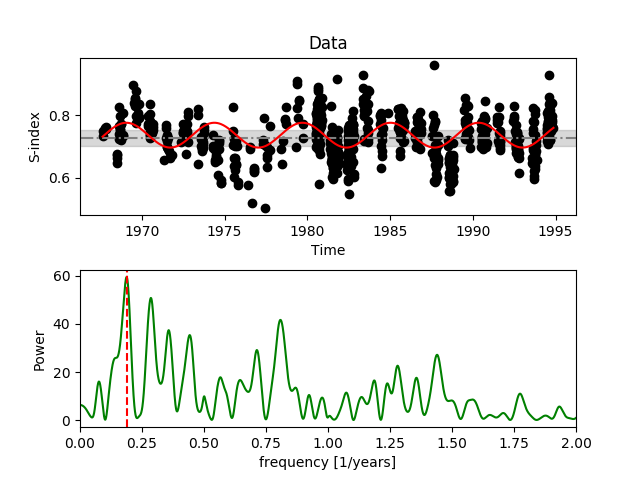}}\\
\subfloat[HD201092]{\includegraphics[scale=0.32]{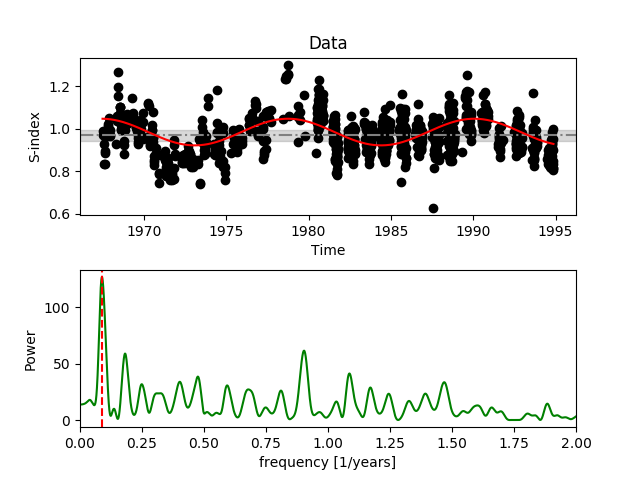}}
\caption{Cool stars (Mount Wilson) with probable solar-like cycles.}
\label{C}
\end{figure*}

\begin{figure*}
\centering
\subfloat[HD20003]{\includegraphics[scale=0.32]{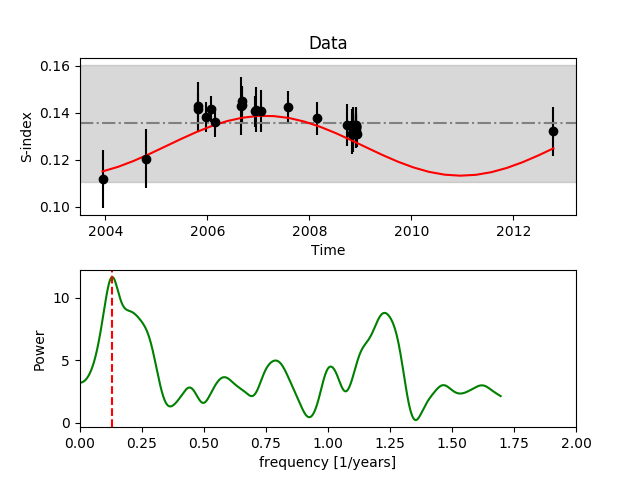}}
\subfloat[HD20619]{\includegraphics[scale=0.32]{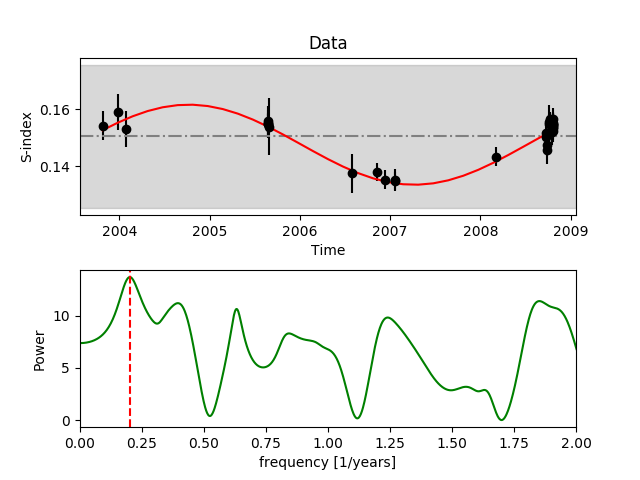}}
\subfloat[HD21693]{\includegraphics[scale=0.32]{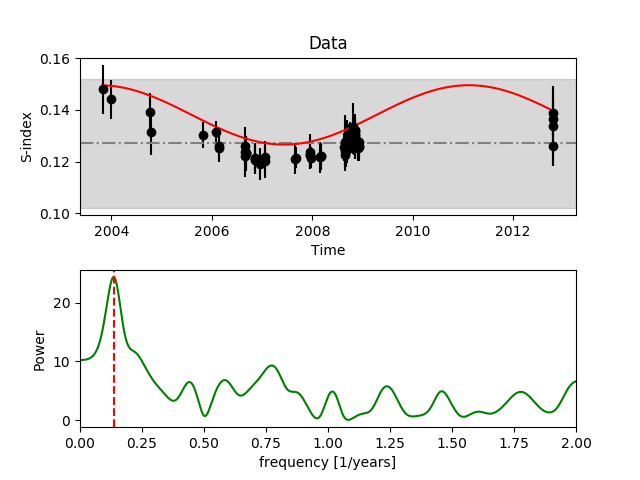}}\\
\subfloat[HD45184]{\includegraphics[scale=0.32]{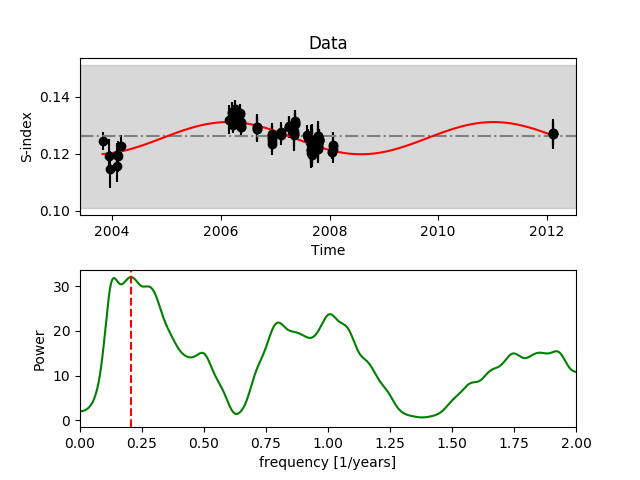}}
\subfloat[HD7199]{\includegraphics[scale=0.32]{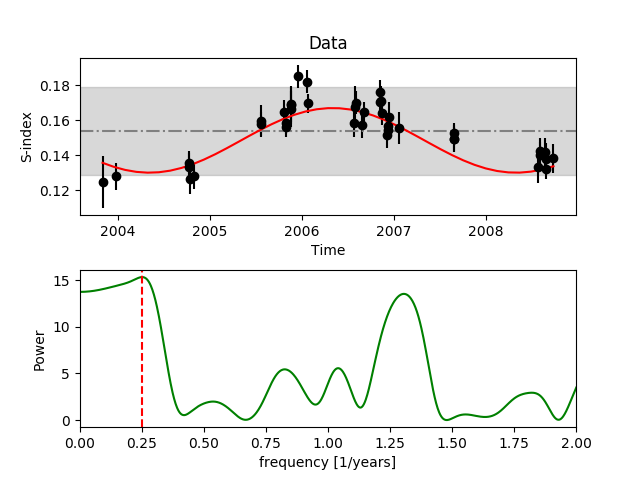}}
\subfloat[HD82516]{\includegraphics[scale=0.32]{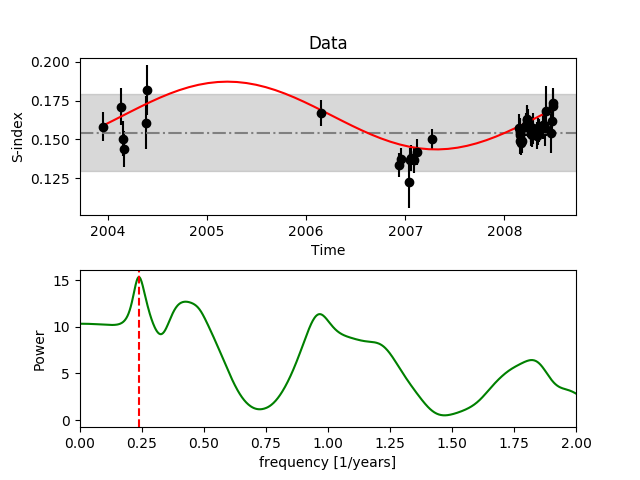}}\\
\subfloat[HD89454]{\includegraphics[scale=0.32]{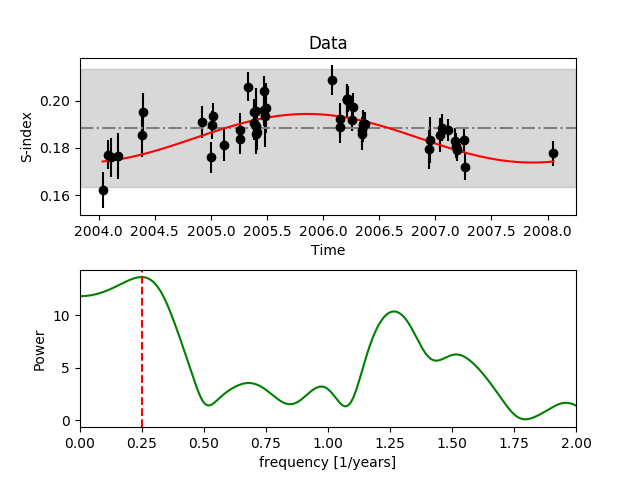}}
\subfloat[HD157830]{\includegraphics[scale=0.32]{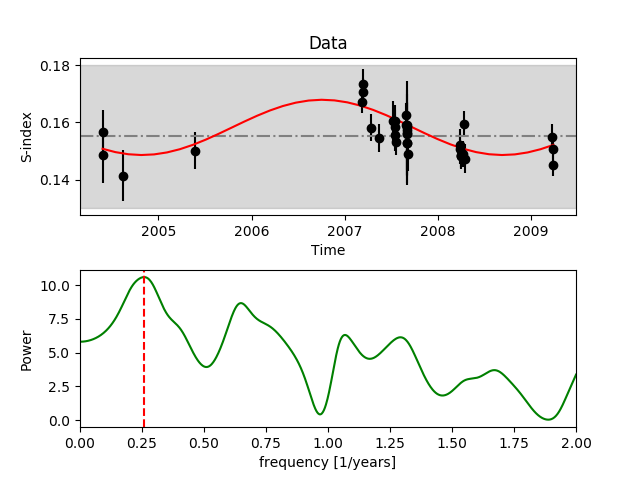}}
\subfloat[HD361]{\includegraphics[scale=0.32]{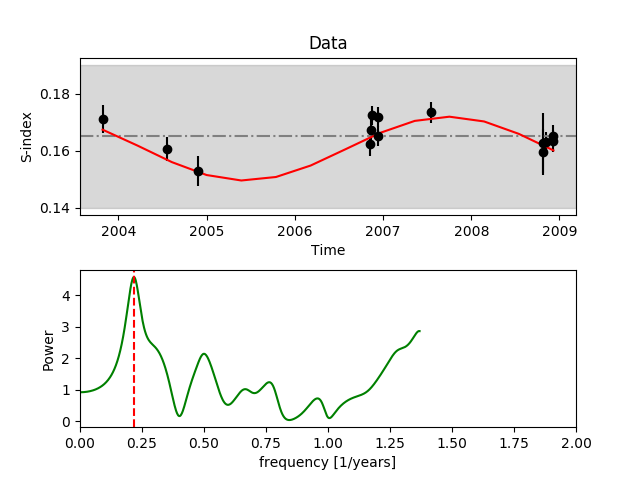}}\\
\subfloat[HD12617]{\includegraphics[scale=0.32]{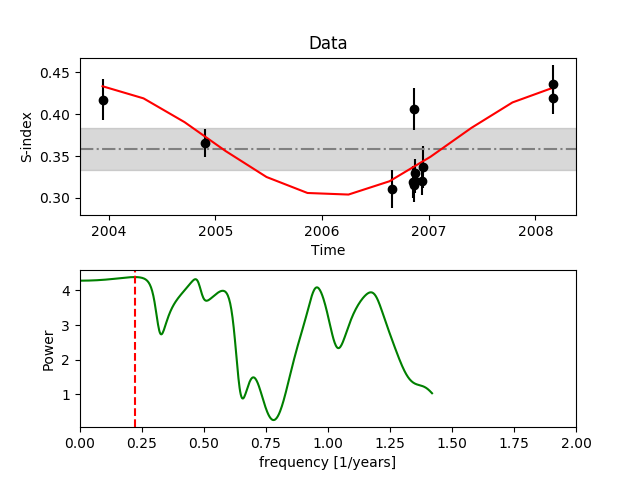}}
\subfloat[HD166724]{\includegraphics[scale=0.32]{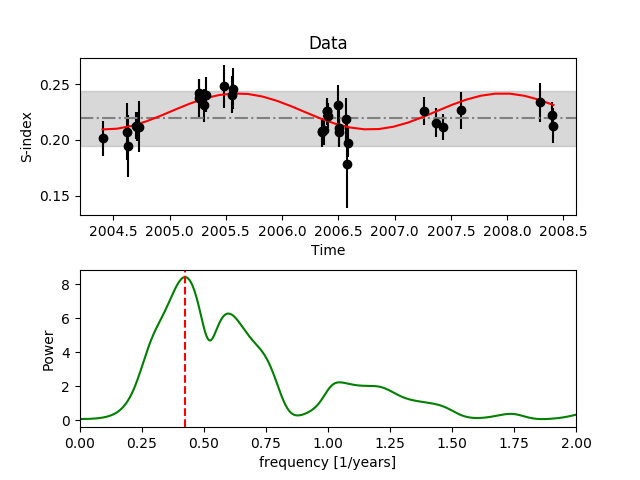}}
\subfloat[HD21749]{\includegraphics[scale=0.32]{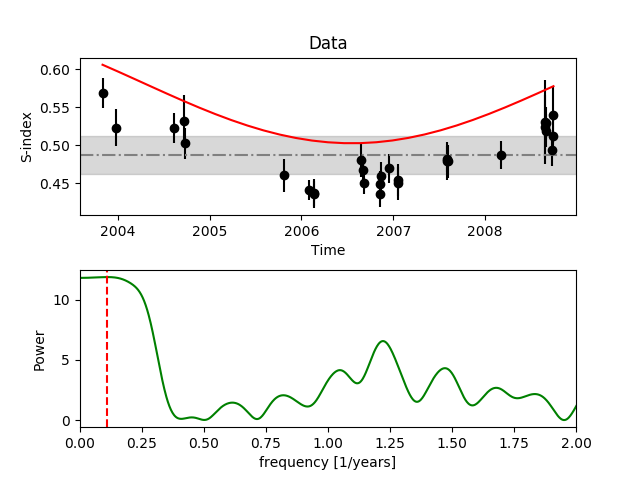}}\\
\subfloat[HD154577]{\includegraphics[scale=0.32]{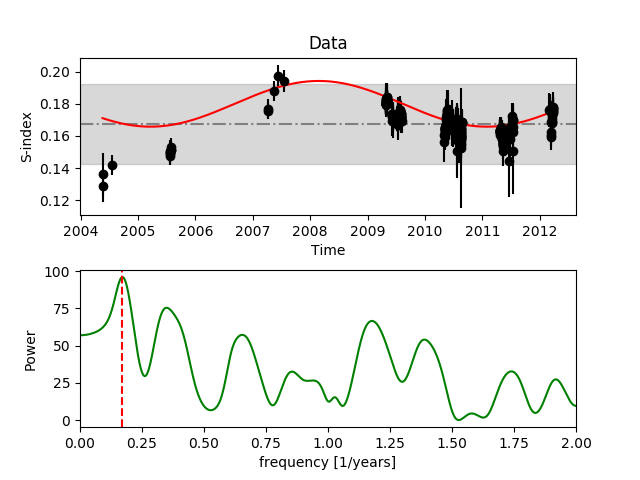}}
\subfloat[HD88742]{\includegraphics[scale=0.32]{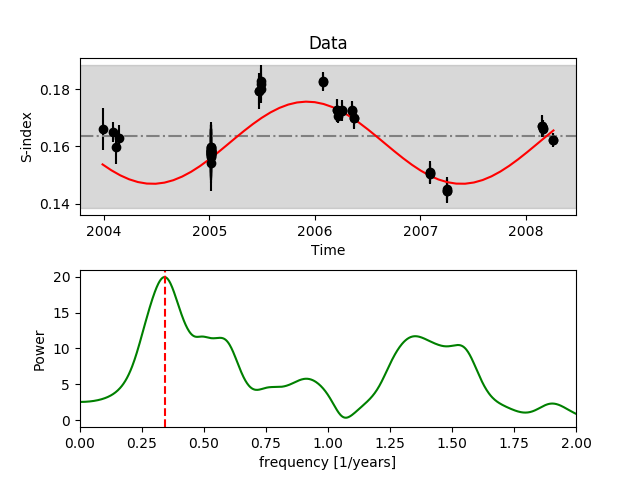}}
\caption{Cool stars (HARPS) with probable solar-like cycles.}
\label{D}
\end{figure*}
\begin{figure*}
\section{Rotation period versus activity cycle for CA stars alone}
\centering
\includegraphics[scale=0.5]{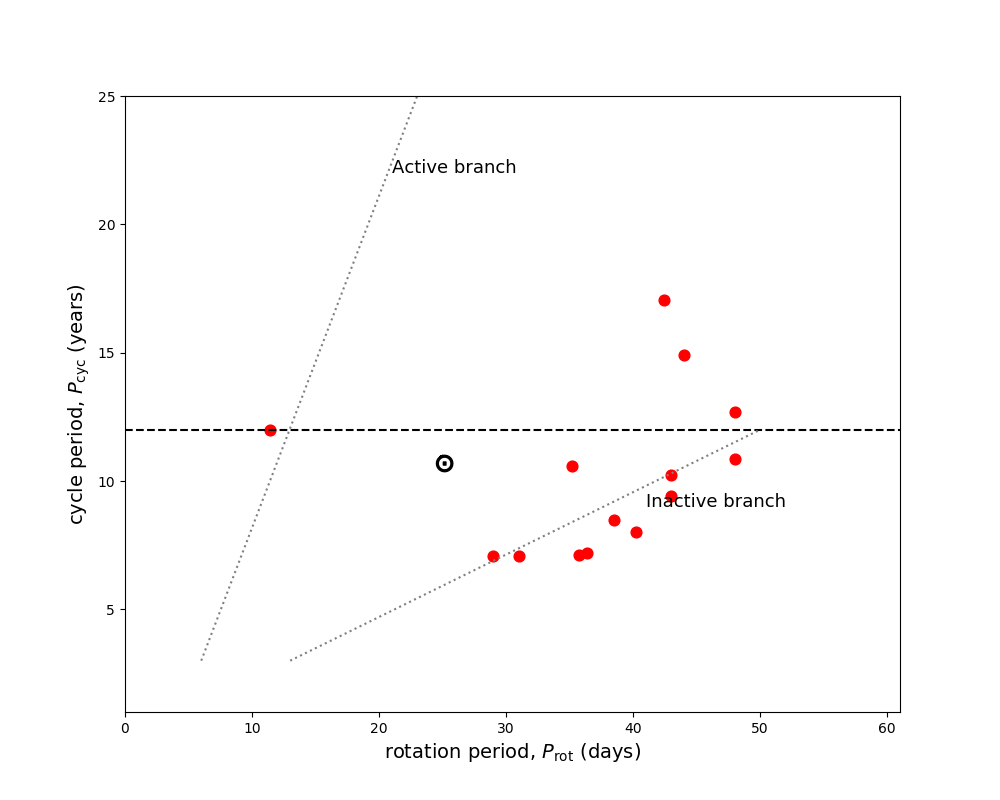}
\caption{Rotation period vs. cycle period where only CA stars are plotted.}
\label{CAfig}
\end{figure*}
\begin{figure*}
\section{Overlap of stars with activity cycles and the Vaughan-Preston gap}
\centering
\includegraphics[scale=0.8]{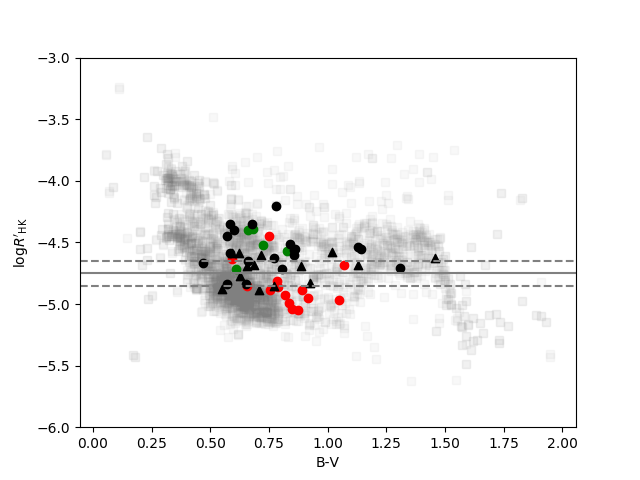}
\caption{Activity vs. $B-V$. The symbols are the same as in 
Figs. 8 and 9. The different colours represent the classification of the cycle periods: red (CA stars), green (CB stars), and 
black (CC stars). The grey squares in the background show the distribution of stars in Fig 3. }
\label{simple}
\end{figure*}
\end{appendix}
\end{document}